\documentclass[natbib]{svjour3}            
\smartqed  
\usepackage{graphicx}
\usepackage{color}
\usepackage{epsfig,amsmath,natbib}
\newcommand{\la}{\raise0.3ex\hbox{$<$}\kern-0.75em{\lower0.65ex\hbox{$\sim$}}}
\newcommand{\ga}{\raise0.3ex\hbox{$>$}\kern-0.75em{\lower0.65ex\hbox{$\sim$}}}

\def\lsim{\;\raise0.3ex\hbox{$<$\kern-0.75em\raise-1.1ex\hbox{$\sim$}}\;}
\def\gsim{\;\raise0.3ex\hbox{$>$\kern-0.75em\raise-1.1ex\hbox{$\sim$}}\;}

\journalname{Space Science Reviews}
\begin{document}

\title{Cosmic rays in galactic and extragalactic
magnetic fields}

\titlerunning{Galactic and extragalactic CRs} 

\author{Felix Aharonian \and Andrei Bykov \and Etienne Parizot \and Vladimir Ptuskin \and Alan Watson}

\authorrunning{Aharonian et al.} 

\institute{Felix Aharonian \at Center for Astroparticle Physics and
Astrophysics, DIAS, Dublin, Ireland and MPIK, Heidelberg, Germany,
\email{Felix.Aharonian@mpi-hd.mpg.de} \and Andrei Bykov \at A.F.Ioffe
Institute for Physics and Technology, 194021 St.Petersburg, Russia also St.Petersburg State Politechnical University \email{byk@astro.ioffe.ru} \and Etienne Parizot \at APC,
Universit\'e Paris Diderot 10, rue Alice Domon et L\'eonie Duquet
75205 Paris Cedex 13 France, \email{parizot@apc.univ-paris7.fr} \and
Vladimir Ptuskin \at Pushkov Institute of Terrestrial Magnetism,
Ionosphere
 and Radio Wave Propagation of the Russian Academy
 of Science (IZMIRAN), Troitsk, Moscow Region 142190, Russia,
 \email{vptuskin@izmiran.ru} \and Alan Watson  \at School of Physics and Astronomy
University of Leeds Leeds, LS2 9JT, UK,
\email{a.a.watson@leeds.ac.uk}}

\date{Received: date / Accepted: date}

\maketitle
\begin{abstract}
We briefly review sources of cosmic rays, their composition and
spectra as well as their propagation in the galactic and
extragalactic magnetic fields, both regular and fluctuating. A
special attention is paid to the recent results of the X-ray and
gamma-ray observations that shed light on the origin of the galactic
cosmic rays and the challenging results of Pierre Auger Observatory
on the ultra high energy cosmic rays. The perspectives of both high
energy astrophysics and cosmic-ray astronomy to identify the sources
of ultra high energy cosmic rays, the mechanisms of particle
acceleration, to measure the intergalactic radiation fields and to
reveal the structure of magnetic fields of very different scales are
outlined. 

\end{abstract}

\section{Introduction}
\label{intro} 
\begin{figure*}
\includegraphics[width=0.70\textwidth]{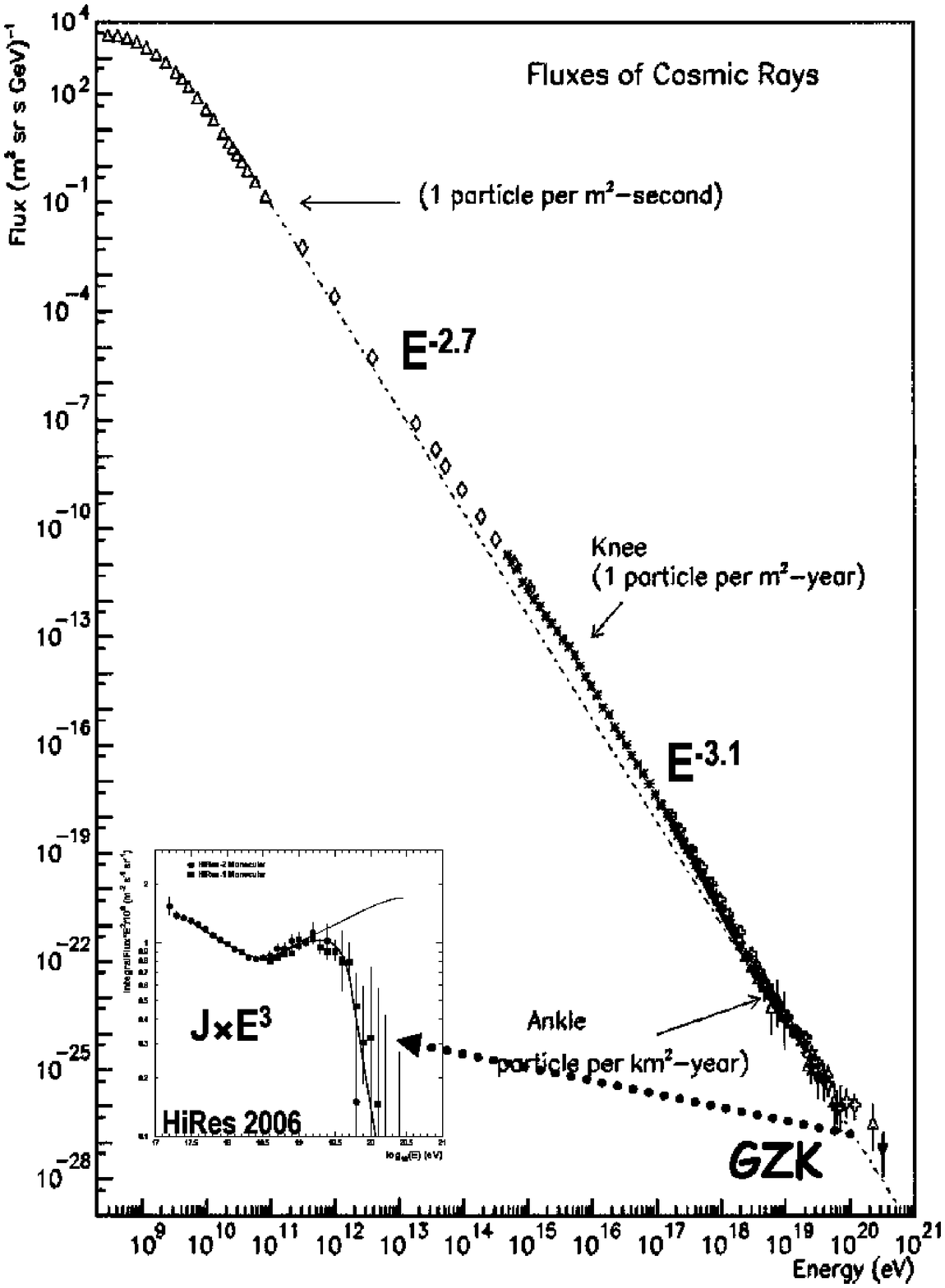}
\caption{All particle spectrum of cosmic rays (after S. Swordy and
from \citet{Abbasi08} -- insert).} \label{spectr}
\end{figure*}

To a first approximation, the all-particle spectrum of
cosmic rays (CR) - ionized atomic nuclei with energies extending
from the MeV to more than $10^{20}$ eV, can be described by a power
law over $11$ decades on particle energy, so that the dependence of
cosmic ray intensity on particle energy is close to $E^{-2.7}$ above
$10$ GeV/nucleon see Figure~\ref{spectr}. Spectra of different CR
species are different however. Closer examination reveals some
structures in the galactic cosmic ray spectrum that includes more
flat low energy part at $E < 10$ GeV/nucleon (mainly due to the
modulation of the Galactic cosmic ray intensity by the solar wind
flow), the knee at about $4\times10^{15}$ eV, the possible second
knee at $\sim 4\times10^{17}$ eV, the ankle at $\sim 4\times
10^{18}$ eV where the spectrum flattens again, and finally the
suppression of flux above $\sim 5\times 10^{19}$ eV. The latter can
be interpreted as the so-called Greisen-Zatsepin-Kuzmin (GZK) cutoff
\citep{Gr,ZatsK} caused by the interaction of the ultra-high-energy
cosmic rays (UHECRs) with the cosmological microwave background
(CMB). Therefore, the suppression of the UHECR flux, taken together
with the detection of an anisotropy in their arrival direction
\citep{AugerScience07} and the absence of any anisotropy associated
with galactic structure is often taken as evidence of the
extragalactic origin of these particles. However, it should be noted
that the reported cutoff at $10^{20}  \rm eV$ could also be related
to the limited  capability of even extreme cosmic accelerators to
boost the energy of individual particles up to $10^{20}  \rm  eV$.
In any case, it is widely believed that the highest energy cosmic
rays above $10^{19} \rm eV$ are of extragalactic origin \citep[see
however][]{Blasi+00,Calvez+10}.

On the other hand, the energy region below $10^{15}$ eV is dominated
by particles produced by Galactic sources. A strong argument for
this is provided by gamma-ray observations of the Galactic disk, and
the comparison with the corresponding observations in the Magellanic
clouds. Furthermore, it is likely that even low-intensity magnetic
fields in the intergalactic space prevent low-energy CRs from
reaching us from distant galaxies on cosmological timescales.
Therefore, a transition from galactic to extragalactic cosmic rays
is expected to occur somewhere between $10^{17}$~eV and
$10^{19}$~eV, while it is tempting to associate the flattening of
the spectrum at the so-called ankle in the energy spectrum as
marking the transition region. However, different interpretations
have been proposed, with a transition at lower energy
\citep{Bere+04,Lemoine05,Aloisio+07}, and this important part of
high-energy CR phenomenology is still much
debated~\citep{Allard+07,Berez09}. It is interesting to note that
this question is strongly related to that of the source composition
of extragalactic CRs \citep{Allard+05}, and of course, the energy
scale of the transition depends on the capacity of the Galaxy to
confine cosmic rays up to high energies, and thus on the intensity
and structure of the Galactic magnetic field.

The spectra of both Galactic and extragalactic cosmic rays are
shaped by two basic processes - the acceleration in the sources and
the subsequent propagation in cosmic magnetic fields and radiation
fields. We concentrate here only on the CR propagation and
acceleration while the extensive review of the galactic and
extragalactic magnetic fields is presented in the other chapters of
the volume \citep[see also][and the references
therein]{ferriere01,beck08,kz08}.

\section{Potential sources}
If one uses the cosmic ray energy requirements and the nonthermal
radiation as a guideline, then the most powerful accelerators of
relativistic particles in the Galaxy should be supernovae and
supernova remnants, pulsars, neutron stars in close binary
systems, winds of young massive stars, and possibly gamma-ray bursts or hypernov\ae. The total power
$L_{\mathrm{cr}}$ needed to maintain the observed energy density
of cosmic rays is estimated as $10^{41}$ erg/s. For the
acceleration by a supernovae, this estimate implies the release of
energy in the form of cosmic rays of approximately $1.5\times
10^{50}$ erg per supernova if the supernova rate in the Galaxy is
1 every 50 years. This value comes to about 10--20\% of the kinetic
energy of the ejecta, which is compatible with the expectations of
the theory of diffusive shock acceleration. This assumes the
acceleration of cosmic rays by the outward propagating shock,
which results from the supernova explosion and propagates in the
interstellar medium or in the wind of the progenitor star.

The rotational energy of a young pulsar with period $P$ that remains
after the supernova explosion is estimated to be
$2\times10^{50}$($10$ \textrm{ms}$/P$)$^{2}$ erg. It is an
additional energy reservoir for particle acceleration and in
particular it could be the source of very high energy
electron-positron pairs. The production of relativistic positrons by
pulsars may explain their presence in cosmic rays, see e.g.
\citet{Profu08}. A general dimensional analysis by
\citet{hillas84,hillas05} provided a useful tool to estimate the
maximal CR energies that can be achieved by electromagnetic
acceleration of charged particles in a source of given size and
magnetic field magnitude. It follows that the highest energy CR
particles of energy $\gsim$ 10$^{20}$ eV  can occur either in the
compact sources with very large magnetic fields or in the most
extended cosmological objects, though the particle energy losses may
affect the maximal CR energy estimations.


\subsection{Supernova remnants as CR accelerators and gamma-ray astronomy}

The observations of non-thermal radio emission of supernova remnants
have established the presence of effective acceleration of cosmic
ray electrons in these objects, e.g. \cite{Lozinsk92,druryea01}.
Shell-type supernova remnants exhibit a broad range of spectral
indices, centered roughly on $\alpha=-0.5$. This implies a power-law
distribution of electrons with an average index
$\gamma=2\alpha+1\approx 2.0$, but it varies in a broad range from
about $1.4$ to $2.6$. The analysis of the synchrotron emission in
the young supernova remnant Cas A \citep{JonesCas02} showed the
presence of electrons with energies up to $200$ GeV with a magnetic
field strength of about $500$ $\mu$G. The interpretation of
nonthermal radio emission from external galaxies proved that
supernova remnants are the sites of acceleration of relativistic
electrons with an efficiency similar to the one needed to provide
the observed intensity of Galactic cosmic-ray electrons
\citep{Duric95}.

In the case of young  SNRs, the spectra of nonthermal  synchrotron
radiation  continues into the X-ray region. In  particular,  bright
nonthermal  X-ray structures  have been  observed from SN~1006,
RX~J1713.7-3946, RCW~86, and Cas~A.
The only viable explanation of this emission is the synchrotron radiation of
ultrarelativistic electrons accelerated to energies as high
as 100 TeV \citep[for a review, see][]{Vink03,Reynolds08}.
Inverse Compton scattering of background photons by these
electrons, together with  gamma rays  produced  via $\pi^{0}$
decays are  two major  mechanisms of  gamma-ray  production
related to the forward and reverse shocks.

It is generally  believed that gamma-ray astronomy has  a key role
to play in  solving the problem of origin of  galactic cosmic
rays. The  basic concept
 is simple and concerns both  the acceleration and propagation
aspects of  cosmic rays.  Namely,  while the localized  gamma-ray
sources exhibit  the sites of production/acceleration of  cosmic
rays, the angular and spectral distributions  of the diffuse
galactic  gamma-ray emission provide information about the
character of propagation of relativistic particles in galactic
magnetic fields. The crucial outcome of current efforts involving
both the ground-based and space-borne gamma-ray detectors    is
hoped to be  a decisive test  of  the popular hypothesis that
supernova  remnants (SNRs)  are responsible for the bulk of the
observed  cosmic ray  flux. Three factors,  (i) high efficiency of
transformation of the available kinetic energy of bulk motion into
nonthermal particles,  (ii) the  predicted, within the DSA
paradigm, hard  energy spectra of protons  extending to multi-TeV
energies and   (iii) relatively high density of the ambient  gas -
make  the young supernova remnants  viable sources of  gamma-rays
resulting from production and prompt decay of  secondary
$\pi^0$-mesons.  Thus,  the  most straightforward  test of
acceleration of proton and nuclei in  SNRs  can be performed via
search for  $\pi^0$-decay  gamma-rays  -   either directly from
shells  of   young SNRs   \citep{DAV94}  or from nearby clouds
interacting with  an expanding SNR shell \citep{ADV94}.

One may argue  that the TeV gamma-ray domain is the best energy
band to explore this possibility - from the point of view of  the
superior performance of the detection  technique and because of
the key information about particle acceleration  carried by TeV
gamma-rays.  On the other hand, TeV gamma-rays are expected only
from young SNRs, when the shock speed can be as high as 3000 km/s
- a key condition for acceleration of protons to energies 100 TeV
and beyond. With the age, typically $t \geq 2000$ yr, the shock
speed and the magnetic field in the expanding shell are decreased
to the point when the particle acceleration significantly  slows
down. Moreover,  the highest energy particles accelerated at the
early epochs, effectively leave the remnant.  This results in
dramatic reduction of gamma-ray emissivity at TeV energies.
Meanwhile the production of GeV gamma-rays  continues.  Thus
 we expect  MeV and GeV gamma-rays not  only from young, but, to a large extent, also from mid-age
 SNRs.   In this regard we expect  significantly  more GeV emitting SNRs compared to the TeV emitting SNRs.
 Unfortunately, due to relatively low gamma-ray fluxes, as well as very high gamma-ray background in
 the crowded regions of the galactic disk, the detection of GeV gamma-rays from SNRs is quite difficult.
  In the case of mid-age  SNRs we expect also  "delayed" gamma-ray emission  produced by cosmic rays which have left the source
 and interact with nearby massive gas molecular clouds. The spectrum of cosmic rays, and, consequently, gamma-rays produced in
 p-p interactions depends on several factors, including the spectrum of protons that escaped  the  supernova remnant, the character of diffusion,
 the age of the source and  the distance between the source and cloud.  In this regard one should expect gamma-ray emission  with GeV-to-TeV
 flux ratio which can vary significantly from a source to source.

Over the last several years   TeV gamma-ray emission   has ben
reported   from  several  SNRs. Presently   six   young shell type
SNRs -  Cas A,   RX J1713.7-3946, RX~J0852-4622 (Vela Jr), RCW~86,
SN~1006, Tycho  - are identified as TeV gamma-ray emitters.
Several others, in particular  IC~433, W28 and  CTB~37B are, most
likely,  SNR/molecular-cloud  interacting  systems. Also, several
galactic  TeV gamma-ray sources spatially coincide with  the
so-called composite supernova remnants,  objects  with
characteristic features  of both  standard shell-type SNRs  and
Pulsar Wind Nebulae \citep[see e.g.][]{ABKS08}.

\begin{figure}[h]
\includegraphics[width=0.95\linewidth]{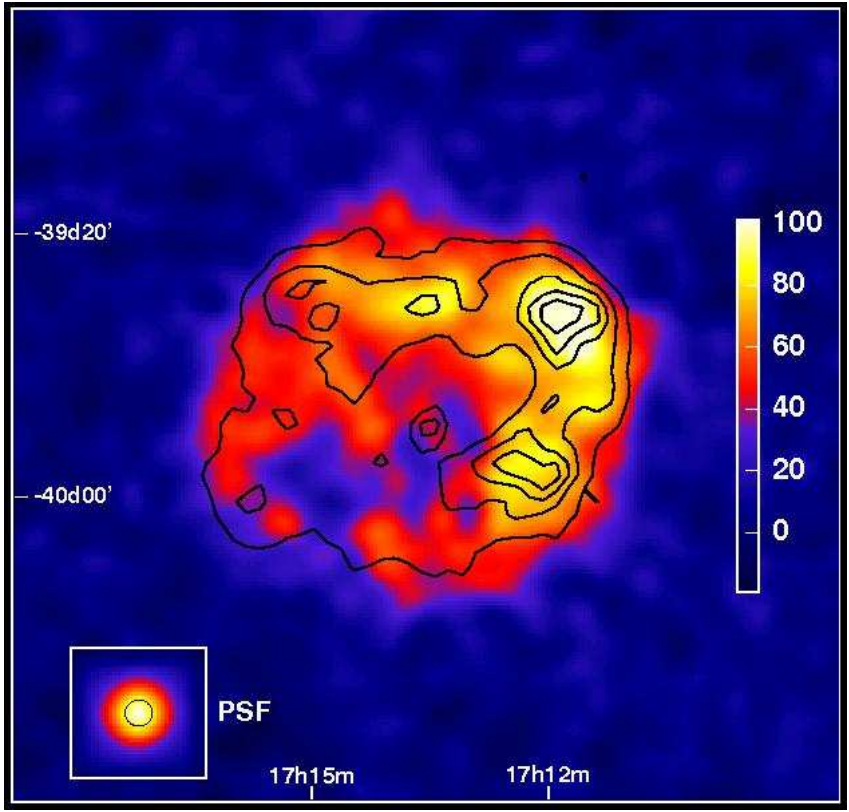}
\caption{The gamma-ray image of RX~J1713.7-3946 obtained with the
H.E.S.S. telescope array \citep{HESS_RX1713_3}. In the {\it
left-hand side}, the overlaid light-gray contours illustrate the
significance of the different gamma-ray features at $8$, $18$, and
$24\sigma$ levels. In the {\it right-hand side},  the black-lines
show the X-ray contours.} \label{RXJ_image}
\end{figure}

In Fig.\ref{RXJ_image} we  shown the TeV and X-ray images of the
strongest and best studied gamma-ray emitting SNR - RX
J1713.7-3946.

The  age of this  SNR RX~J1713.7$-$3946  at a distance of about 1~
kpc is estimated  between 1 to 3 thousand years.  Therefore, it
can be formally teated as a representative of young galactic SNRs
like SN~1006, Vela~Junior, Tycho, Kepler, Cas A. However,
RX~J1713.7-3946  seems to be a unique object with  quite  unusual
characteristics.  This circumstance makes the identification of
the gamma-ray mechanism, and hence the nature of the parent
particles a nontrivial problem. Meanwhile the question of the
origin of gamma-rays has a fundamental  implication related to the
role of  SNRs to the production of galactic cosmic rays.

Remarkably,  both the hadronic (gamma-rays produced in pp
interactions) and leptonic models (gamma-ray produced via inverse
Compton scattering) provide, with more or less success, but yet
acceptable fits to the measured gamma-ray spectra and explain also
the spatial distributions  of gamma-rays and X-rays
\cite{ZirAhar10}. So the test of these models relies essentially
on multiwavelength data.

One of the most  puzzling  features of  RX~J1713.7-3946  is the lack
of thermal  X-ray emission which  is a serious  argument against the
hadronic models of  of this source. The tight upper limit on the
thermal X-ray flux of   RX~J1713.7-3946  is explained by the
supernova explosion  inside a  wind-blown bubble  with a very  low
gas density  $n_{\rm gas}  \ll  1 \ \rm cm^{-3}$. This allows to
suppress the free-free component of thermal emission but not the
X-ray lines \citep{Elisson2010,ZirAhar10}, thus an additional
assumption about the peculiar  composition with reduced content of
heavy elements is required to avoid the conflict with X-ray data
\citep{ZirAhar10}. The hadronic  models  demand  quite large
magnetic field of order of $100 \ \rm \mu G$ or larger. This, for
the given flux of synchrotron radiation,  requires only $\approx
10^{46} \ \rm erg$ in electrons. On the other hand,  the total
energy in protons should be $10^{50}$~erg or  significantly larger
if the background gas density $n \ll 1 \ \rm cm^{-3}$. Thus,   in
the hadronic scenario the electron-to-proton ratio is close to
$K_{\rm ep}=10^{-4}$. This is  a   two orders of magnitude smaller
than the "standard" electron/proton  ratio observed in  local cosmic
rays.

A major challenge for the leptonic models is the demand for a
small magnetic field in the shocked shell of the remnant. Within
the framework of a one-zone model in which the electron
acceleration and gamma-ray emission take place in the same region,
the magnetic field cannot significantly exceed  $10~\mu
\mathrm{G}$, otherwise it leads, for the given X-ray flux, to
significant suppression of IC gamma-rays.

\begin{figure}[t]
\includegraphics[width=7.5cm]{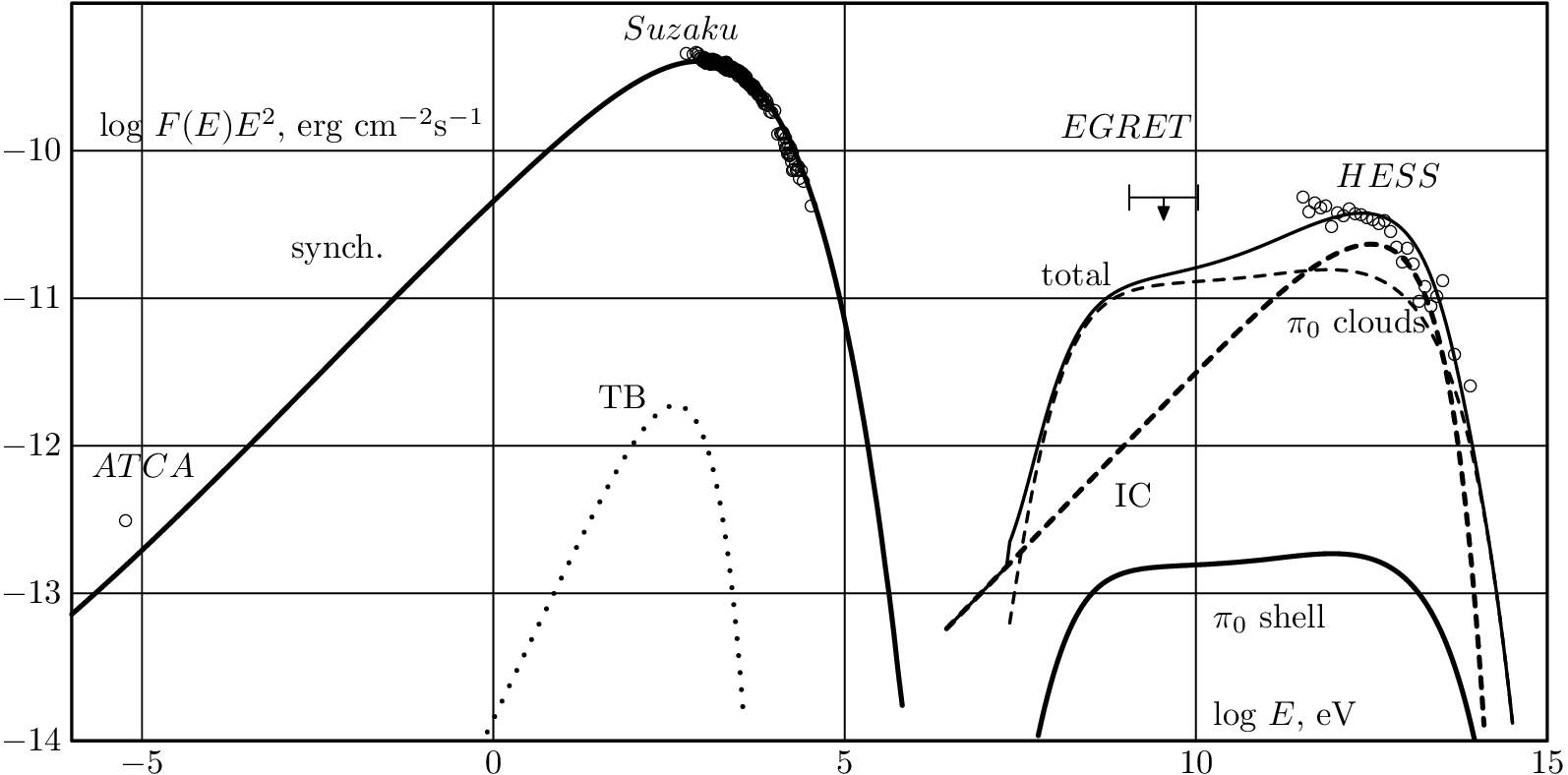}
\caption{Broad-band  emission of RX J1713.7-3946  for the
"composite" scenario of  gamma-rays with non-modified forward and
reverse shocks  and dense clouds in the shell \cite{ZirAhar10}.}
\label{VZ}
\end{figure}

A possible solution to the problems related to the pure
(oversimplified) one-zone leptonic and hadronic models could be
effective acceleration of  electrons and protons in both forward
and reverse shocks with an additional assumption of existence of
dense condensations  embedded in very low density shell.  In this
scenario  the overall VHE gamma-ray emission is dominated by the
IC scattering of electrons, but in addition to the leptonic
component one should expect non-negligible contribution    from
hadronic gamma-rays produced at interactions of protons with dense
gas condensations. Moreover,  in this scenario both the  low
energy, $E \leq 100$~GeV and highest, $E \geq 10$~GeV gamma-rays
should arrive from specific regions occupied by dense gas
condensation (see Fig.\ref{VZ}). The "composite"  scenario allows
a quite relaxed parameter space compared to the "pure"  hadronic
model and leptonic models, and at the same time, provides better
fits to the broad-band spectral energy distribution of the
source.

Independent of the ability of different models to describe the
spectral and morphological features of  gamma-ray  emission of RX
J1713.7-3946, it is obvious that we deal with a  source  that
effectively accelerates   electrons and protons  to energy of  100
TeV and beyond. On the other hand,  this  object has many unique
features,  and in this regard it could be misleading if we treat
this source as a representative of  the whole population of young
SNRs. Therefore, the studies of other TeV emitting  SNRs is
crucial for definite conclusions concerning the role of SNRs in
production of galactic cosmic rays.

The realization of different possible scenarios of particle
acceleration and radiation in general, and the uncertainties in
relevant model parameters like the strength of the magnetic field
and the density of ambient gas, prevent us from unambiguous
conclusions concerning the fraction of the hadronic component in the
flux of observed gamma-rays, and consequently do not allow
model-independent estimates of thew total energy released in
accelerated protons and nuclei.  The best hope in this regard is
related to the detection of high energy neutrinos produced in
interactions of accelerated protons and nuclei with the ambient
plasma. Although the fluxes of neutrinos are quite similar to the
gamma-ray fluxes, because of limited sensitivity of the neutrino
detectors, this extremely important channel of information
unfortunately remains unexplored. Even for the largest neutrino
telescopes like the recently completed \emph{IceCube} and the
planned several km-cube underwater detector in the Mediterranean
Sea, the high energy neutrino signal from RXJ 1713.7-3946 is
expected to be marginal even under the extreme assumption that the
flux of detected TeV gamma-rays is fully consists of photons of
hadronic origin \citep[see e.g.][and the references
therein]{vissaniea11,mba09}.

\subsection{OB-associations, starburst, superbubbles}
  Collective stellar winds and supernovae in clusters and
associations of massive stars in galaxies can be potentially
efficient cosmic ray accelerators \citep[see
e.g.][]{cm82,bf92,axford94,Higdon98,b01,bt01,binns07}.

The young rich open cluster Westerlund 2 is spatially coincident
with TeV photon source HESS J1023-575 \citep{HESSwesterlund2}.  The
reported source extension excludes a single star origin of the
observed VHE emission and argues in favor of the production of
cosmic rays  by collective stellar winds from the ensemble of at
least a dozen of hot massive O-type and a few WR-type stars in the
stellar cluster Westerlund 2 (of an estimated age of about a few
million years).

Recently, TeV gamma-ray emission has be reported from two star-burst
galaxies, NGC 253 and M82, by H.E.S.S. \citep{Hess_starburst09} and
VERITAS \citep{VERITASNature09}, respectively.
Starburst galaxies exhibit in their central regions a highly increased rate of
supernovae that are likely to be efficient producers of cosmic rays. As they
are extended, with radii up to $\sim 100$~pc, the superbubbles (SB) containing dozens of
supernova remnants and winds of massive stars are difficult to
identify in the Galaxy, and are thus currently best studied in the Magellanic Clouds.
Nonthermal X-rays were detected by
\citet{bambaea04} from the shells of the superbubble  30 Dor C in
the LMC. The X-ray observations revealed the SB morphology as a nearly circular
shell with a radius of ~40 pc, which is bright on the northern and
western sides. The non-thermal X-ray shell traces the outer boundary
of the radio shell. These features of thin thermal and non-thermal
X-rays are similar to those of SN 1006, a prototype of a synchrotron
X-ray shell, but the non-thermal component of 30 Dor C is about 10
times brighter than that of SN 1006.
The source is also much older than that of SN
1006, so the particle acceleration time in this superbubble may be
longer than those in normal shell-like supernova remnants.

Nonlinear modeling of CR acceleration in superbubbles by
\citet[][]{b01} predicted a high efficiency of the kinetic energy
conversion to the CRs that can be $\sim$ 20\% and even higher  with
a significant temporal evolution of particle spectra inside the
superbubble. Magnetic field inside the superbubble can be amplified
 up to about 30 $\mu$G providing the maximal energies of CR protons
to be about 10$^8$ GeV and higher for CR nuclei that dominate  the
accelerated CR composition at highest energies in the model
\citep[see][]{bt01}. To model the CR composition in superbubbles a
number of components including the $^{22}$Ne-rich Wolf-Rayet and
dust material must be accounted for \citep[][]{binns07,mde97,edm97}.

\subsection{CR abundance constraints on potential CR sources}

The correction of the cosmic ray composition observed at the
Earth for the effects of nuclear fragmentation in the interstellar medium makes it possible
to determine the initial elemental and isotopic composition of
accelerated particles, to clarify the process of cosmic ray
acceleration and the nature of cosmic ray sources.

The relative abundance of chemical elements in cosmic ray sources is
in general similar to the solar and local galactic abundances but
with some interesting deviations. The popular scenarios which
explain the cosmic ray source composition include the acceleration
of grains together with relatively less abundant volatile ions by
shocks in the interstellar medium \citep{mde97,edm97}, the
acceleration of freshly formed material, particularly grains in
young supernova remnants \citep{Lingenfelt98}, and the acceleration
in hot superbubbles with multiple supernova remnants \citep[see
e.g.][]{binns07}.

The isotopic composition of cosmic rays is now measured for all
stable isotopes for elements from H to Ni. The isotopic composition
of cosmic ray sources is strikingly similar to the composition found
in the solar system \citep{Weidenbeck99}. However, a
well-established anomaly in the isotopic composition of galactic
cosmic rays is the excess of $^{22}$Ne. The ratio
$^{22}$Ne/$^{20}$Ne is enhanced by a factor of $4$ compared with the
solar reference value \citep{DuVernois96}. It can be explained only
by the special conditions of nucleosynthesis. The enhancement of
neutron rich isotopes would be expected in the highly evolved very
massive stars in their Wolf-Rayet stage when their surfaces contain
large excesses of the products of core helium burning, including
$^{22}$Ne \citep{CassePaul82}. An increased cosmic ray ratio C/O by
a factor of $2$ is also in favor of Wolf-Rayet stars.

Using the data from \emph{Cosmic Ray Isotope Spectrometer} aboard
the \emph{ACE} spacecraft \citet{binns07} measured the isotopic
abundances of neon and a number of other species in the galactic
cosmic rays.  The authors found that the three largest deviations of
galactic cosmic ray isotope ratios from solar-system ratios, namely
$^{12}$C/$^{16}$O, $^{22}$Ne/$^{20}$Ne, and $^{58}$Fe/$^{56}$Fe, are
consistent with a model of CR source consisting of about 20\% of
Wolf-Rayet  wind material mixed with about 80\% material of the
solar-system composition. Therefore \citet{binns07} concluded that
OB associations within superbubbles are the likely source of at
least a substantial fraction of GCRs.

The analysis of the $^{59}$Co/$^{56}$Fe abundance ratio also
provides important information. Indeed, the relatively high, close
to the solar value, of this ratio \citep{Mewaldt99} testifies that
the major part of the originally synthesized $^{59}$Ni, which is
stable when it is ionized (i.e. when it is accelerated among the
cosmic rays), has decayed by the K-capture of an orbital electron
into $^{59}$Co before the acceleration started. Thus, the delay
between the synthesis of this material and its acceleration has to
be larger than $10^{5}$ yr severely constraining the models with the
acceleration of freshly ejected matter in SNRs.

The production of cosmic rays with power $L_{cr} = 10^{41}$ erg/s
per normal galaxy can provide a cosmic-ray energy density of $w_{Mg}
= L_{cr} N_g T_H = 3\times 10^{-17}$ erg cm$^{-3}$, where $N_g =
2\times 10^{-2}$ Mpc$^{-3}$ is the number density of galaxies and
$T_{H}=1.4\times10^{10}$ yr is the Hubble time. This is almost $5$
orders of magnitude smaller than the energy density of cosmic rays
observed in the solar system. The contribution of active galaxies
can reduce this discrepancy by only a factor of $10$. All this is in
agreement with the picture described in the Introduction, where the
main part of the cosmic rays we observe at the Earth is produced by
galactic sources (e.g. supernova remnants) and slowly diffuse in
galactic magnetic fields out of the Galaxy to the intergalactic
space, where the cosmic ray density is much smaller. The efficiency
of the confinement of energetic particles is decreasing with energy,
and because of that the difference between the cosmic ray densities
inside and outside the Galaxy is also decreasing. The densities
become equal at some energy between $10^{17}$ and $10^{19}$ eV and
the extragalactic sources dominate at higher energies. The list of
potential extragalactic sources includes in particular the active
galactic nuclei, the progenitors of gamma-ray bursts, the accretion
shocks of the size of galactic clusters and larger \citep[see][for a
review]{Torres04,Sig09}.
These objects are probably able to provide the required power
$q_{Mg}=3\times 10^{36}$ erg s$^{-1}$ Mpc$^{-3}$ in cosmic rays with
energies $>10^{19.5}$ eV. This estimate of $q_{Mg}$ is based on the
cosmic ray flux determined using the Auger Observatory
\citep{Auger10} and it takes into account the energy loss time due
to the GZK effect. It is notable that the entire population of
supernovae in the Universe also meet the required power condition
\citep{pariz06}, but as far as we know the highest energy
particles
can not be accelerated in such sources. The origin of the highest
energy CRs in extragalactic sources remains a mystery, whose solution will require enhanced
detections capabilities (because of the very low fluxes involved, see below), but will have
a strong impact on the whole field of high-energy astrophysics.

\section{Transport in the Galaxy}
\label{galtrans}
The interaction of relativistic charged particles with
galactic magnetic fields explains the high isotropy and relatively
large confinement time of the cosmic rays in the Galaxy. It is
accepted that the diffusive approximation gives an adequate
description of the cosmic ray propagation in the interstellar medium
at energies up to $\sim 10^{17}$ eV. The diffusion model forms a
basis for interpretation of the cosmic ray data and the related
radio-astronomical, X-ray and gamma-ray observations.

\subsection{Empirical model of cosmic ray propagation}\label{dif}
The procedure to model the cosmic ray diffusion in the Galaxy can be
summarized as follows. One must first specify the cosmic ray
sources, define the shape of the cosmic ray halo and the conditions
at its boundaries (it is usually assumed that energetic particles
exit freely into the intergalactic space where the cosmic ray
density is negligible). The basic diffusion-convection equations for
different cosmic ray species describe the diffusion of relativistic
charged particles and, if needed, their convective transport in the
models including a galactic wind. The equations should incorporate
all possible energy loss and gain processes in the interstellar
medium, nuclear fragmentation, and the radioactive decay of unstable
nuclei. One can then calculate the distribution functions of
protons, electrons and the different types of nuclei. The transport
coefficients of cosmic rays (diffusion coefficient and convection
velocity), the properties of the cosmic ray sources (total power,
energy spectra of the different components, elemental and isotopic
composition), and the size of the confinement region of cosmic rays
in the Galaxy can be inferred by fitting all the available data on
cosmic rays. Hundreds of stable and radioactive isotopes are
included in the calculations of nuclear fragmentation and the
transformation of the energetic nuclei as they interact with the
interstellar gas. The most advanced code developed for the numerical
modeling of cosmic ray propagation in the Galaxy is the GALPROP code
which uses a Crank-Nicholson implicit second-order scheme
\citep{StrMosk98,Str07}. It incorporates as much realistic
astrophysical input as possible, together with the latest
theoretical development, and numerically solves the transport
diffusion-convection equations for all cosmic-ray species.

One of the key channels of information about cosmic-ray
propagation is the abundance of the secondary energetic nuclei Li, Be,
B , Sc, V, Ti, $^{2}$H, $^{3}$He and others produced as a result
of the spallation of heavier primary nuclei interacting with the
interstellar gas. The observed flux ratios between secondary and
primary nuclei, for example the boron to carbon ratio, show a
maximum at about $E = 1$ GeV/nucleon (e.g.,
\citep{Eng90,Yanas01,CREAM08}.
The mass distribution of secondary nuclei can be understood in terms
of their propagation through about 10~g/cm$^{2}$ of interstellar
material. This quantity, referred to as the escape length, can be
approximately expressed through the parameters of the plain
diffusion model with a flat cosmic ray halo of total thickness
$2H\approx 8$ kpc as $X \approx v\mu_{g}H/2D$, where $v$ is the
particle velocity, $\mu_{g}\approx 2.4$ mg/cm$^{2}$ is the surface
gas density of galactic disk, and $D$ is the cosmic ray diffusion
coefficient. This allows an estimate of the diffusion coefficient:
$D \sim 3\times 10^{28}$ cm$^{2}$/s at $E = 1$ GeV/nucleon.

The primary-to-secondary ratios decrease at higher energies (at
least up to a few hundred GeV/nucleon where the measurements exist)
as well as at lower energies. The understanding of this behavior is
provided by the kinetic theory of particle interaction with the
interstellar magnetic fields.

\subsection{Kinetic theory of cosmic ray diffusion in galactic magnetic fields}

The theory of energetic particle transport in the galactic magnetic
fields is constructed in much the same way as in the better studied
case of particle transport in the interplanetary magnetic fields.
The detailed treatment of cosmic ray diffusion in magnetic fields
can be found in monographs \citep{Top85,Ber90,Schlick02}
\citep[see also][]{Ptus06}.


On the ``microscopic level'' the spatial and momentum diffusion of
cosmic rays results from the particle scattering on random MHD waves
and discontinuities. The effective "collision integral" for
energetic charged particles moving in small-amplitude random fields
$\delta B\ll B$ can be taken from the standard quasi-linear theory
of plasma turbulence, see e.g. \citet{Ken66}. The wave-particle
interaction is of resonant character and depends on the particle
gyroradius, which is equal to $r_{g}=pc/ZeB=10^{12} R_{GV}$ cm in a
typical interstellar magnetic field $B=3$ $\mu$G (here $R_{GV}$ is
the particle magnetic rigidity in GV and $\mu$ is the cosine of
particle pitch angle). The particle is predominantly scattered by
the irregularities of magnetic field $\delta B$ that have the
projection of the wave vector on the magnetic field direction equal
to $k_{\parallel}=\pm s/\left( r_{\mathrm{g}}\mu\right)$ . The
integers $s=0,1,2...$ correspond to the cyclotron resonances of
different orders. The efficiency of particle scattering depends on
the polarization of the waves and on their distribution in
$\mathbf{k}$-space. The first-order resonance $s=1$\ is the most
important for the isotropic and also for the one-dimensional
distribution of random MHD waves along the average magnetic field,
$\mathbf{k}\parallel\mathbf{B}$. In some cases -- for calculation of
scattering at small $\mu$ and for calculation of perpendicular
diffusion -- the broadening of resonances and magnetic mirroring
effects should be taken into account. Locally, the cosmic ray
diffusion is anisotropic and goes along the local magnetic field.
The isotropization of diffusion is accounted for by the presence of
strong large-scale ($\sim$ 100 pc) fluctuations of the galactic
magnetic field. The problem of calculation of the average diffusion
tensor is not trivial even in the case of relatively weak large
scale random fields, since the field is almost static and the
strictly one-dimensional diffusion along the magnetic field lines
does not lead to non-zero diffusion perpendicular to $\mathbf{B}$,
see \citet{Cas01,Webb09}.

The following simple equations for the diffusion coefficient are
useful for estimates:
\begin{equation}
D=vr_{\mathrm{g}}B^{2}/\left[  12\pi
k_{\mathrm{res}}W(k_{\mathrm{res} })\right]
=\frac{vr_{\mathrm{g}}^{a}}{3(1-a)k_{\mathrm{L}}^{1-a}}\frac{B^{2}}{\delta
B_{\mathrm{L}}^{2}},
\label{eqD}
\end{equation}
here $k_{\mathrm{res}}=1/r_{g}$ is the resonant wave number, and
$W(k)$ is the spectral energy density of waves normalized as $\int
dkW(k)=\delta B^{2}/4\pi$. The random field at the resonance scale
is assumed to be weak, $\delta B_{\mathrm{res}}\ll B$. The principle
wave number of the turbulence is $k_{L}$ and the amplitude of random
field at this scale is $\delta B_{L}$. The second equation in ($1$)
is valid for particles with $r_{\mathrm{g}}<k_{\mathrm{L}}^{-1}$\
under the assumption of a power law spectrum of turbulence
$W(k)\propto1/k^{2-a},$ $k>k_{_{\mathrm{L}}}$. The diffusion
coefficient has then a power law scaling on momentum $D\propto
v(p/Z)^a$.

Eq.~(\ref{eqD}) can be used to describe the turbulence which
consists of random isotropic distribution of Alfv\'en and fast
magnetosonic waves and it gives correct order of magnitude estimates
for the wave distribution concentrated around the direction of
average magnetic field. It should be pointed out however that the
isotropization of the diffusion tensor does not occur in the case of
a pure parallel wave propagation ($\mathbf{k}\parallel\mathbf{B}$).
Another special case is  $2$D turbulence with perpendicular
propagation of waves ($\mathbf{k}\perp\mathbf{B}$).  In this case,
the scattering occurs only for magnetosonic waves through the
resonance $s=0$ which leads to a very large diffusion coefficient,
about a factor $(v/V_{\mathrm{a}})^{2}$ larger than given by
eq.~(\ref{eqD}).

The diffusion in momentum is approximately described by the
equation:
\begin{equation}
D_{pp}=p^{2}V_{\mathrm{a}}^{2}/\left(9D\right).
\label{eqDp}
\end{equation}
Here equal intensities of waves moving along the magnetic field in
opposite directions are assumed. The  momentum diffusion is vanished
($D_{pp}$ = 0) if all waves move in one direction along the magnetic
field with the same phase velocity. The flow of waves carry
particles along the field in this case and the convection with
velocity determined by the phase velocity of waves appears in the
transport equation for cosmic rays, see Skilling (1975a), Berezinsky
et al. (1990) for detail. Clearly the cosmic ray convection also
arises as a result of the large scale motion of the interstellar gas
with frozen magnetic field.

The random component of interstellar magnetic field with an
extended spectrum of inhomogeneities can give the resonant
particle scattering and spatial diffusion of cosmic rays.
Information on the interstellar turbulence spectrum have been
obtained from radio scintillation and refraction observations
(sensitive to fluctuations of thermal electron density),
measurements of the differential Faraday rotation angles from
distant sources (mainly produced by fluctuations in the
interstellar magnetic field), and the observations of random
motions in the interstellar gas. These data are consistent with
the assumption that a single close-to-Kolmogorov spectrum extends
from scales $10^{8}$ to $3\times 10^{20}$ cm, see \citet{Arm95}
and references therein. The Kolmogorov spectrum is of the form
$W(k)\propto k^{-5/3}$. Other types of spectra frequently used to
describe the interstellar turbulence are $W(k)\propto k^{-2}$ for
the shock-dominated turbulence, see e.g. the model by
\citet{Bykov}, and spectrum $W(k)\propto k^{-3/2}$ suggested by
\citet{Iro63} and \citet{Kra65} in their phenomenological theory
of MHD turbulence. Comprehensive reviews of  MHD turbulence with
application to the interstellar conditions have been given by
\citet{Zho04}, \citet{Elm04}, and \citet{Sca04}.

The estimate based on the empirical value of the diffusion
coefficient for GeV particles (see section \ref{dif}) gives
the level of turbulence at the principal scale $k_{\mathrm{L}}%
=10^{-21}$ cm$^{-1}$ of the order $\delta B_{\mathrm{tot}}/B\sim$\
$0.2$ for an Iroshnikov-Kraichnan spectrum $W(k)\propto k^{-3/2}$
($a=1/2$) and $\delta B_{\mathrm{tot}}/B\sim1$ for a
Kolmogorov-type spectrum $W(k)\propto k^{-5/3}$ ($a=1/3$). At the
same time, the data on Faraday rotation angles favor the
Kolmogorov spectrum with $\delta B_{\mathrm{tot}}/B\sim1$ and
$k_{\mathrm{L}}=10^{-21}$ cm$^{-1}$. The cascades of Alfv\'en waves
(with the scaling $k^{-5/3}$) and the fast magnetosonic waves
($k^{-3/2}$) are independent and may coexist in the \citet{Gol95}
model of MHD turbulence. The amplitude of Alfv\'en wave cascade may
dominate at the principle scale.

There are two physical explanations of the observed
secondary-to-primary nuclei ratios in the diffusion model with a
static cosmic ray halo.

\begin{figure*}
\includegraphics[width=1.00\textwidth, angle=-90]{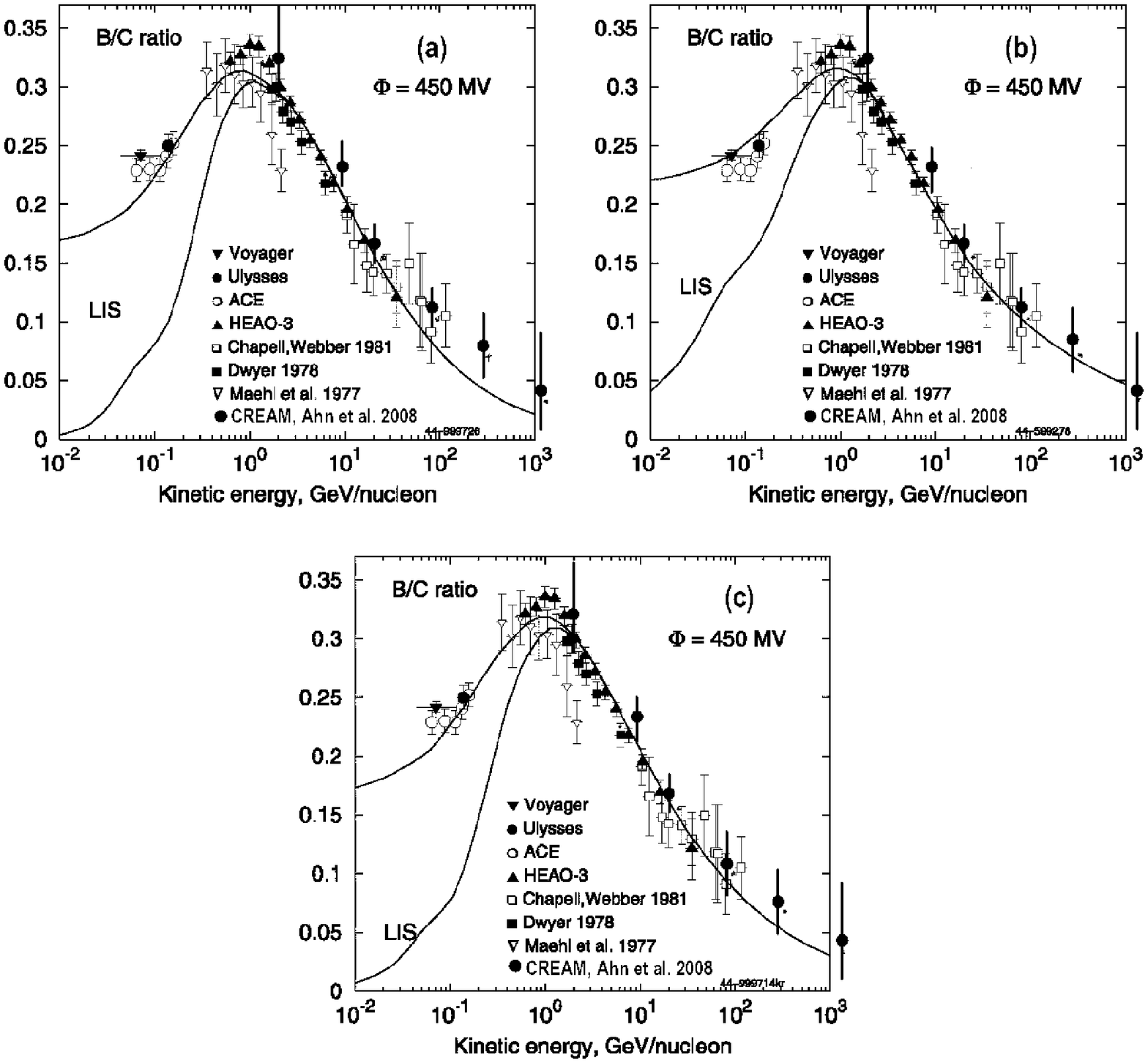}
\caption{B/C ratio as calculated in plain diffusion model with
unphysical break of the diffusion coefficient(a), reacceleration
model (a), and diffusive reacceleration with damping model (c).
Lower curve -- local interstellar spectra, upper -- modulated (with
modulation parameter $\Phi=450$ MV). Data below 200 MeV/nucleon: ACE
\citep{davis}, Ulysses \citep*{ulysses_bc}, Voyager
\citep*{voyager}; high energy data: HEAO-3 \citep{Eng90}, CREAM
\citep{CREAM08} for other references see
\citet{StephensStreitmatter98}.} \label{B_C}
\end{figure*}

The first explanation \citep{Simon,SeoPt} refers to the diffusion
model with distributed reacceleration of cosmic rays by the
interstellar MHD turbulence which scatters particles and provides
their spatial diffusion. The Kolmogorov-type spectrum is assumed
that leads to the rising with rigidity of the diffusion coefficient
$D\propto v(p/Z)^{1/3}$. For typical value of Alfv\'en velocity
$V_{a} \sim 30$ km s$^{-1}$, the reacceleration is not essential for
nuclei with energies $E > 40$ GeV/nucleon and the abundance of
secondary nuclei is a decreasing function of rigidity. The impact of
reacceleration on spectra of primary and secondary nuclei  becomes
stronger at smaller energies so that the characteristic time of
distributed acceleration in the Galaxy becomes equal to the time of
diffusion from the Galaxy at about $1$ GeV/nucleon. In consequence
the pronounced peak in the secondary-to-primary ratio arises, see
Figure~(\ref{B_C}). The asymptotic behavior of the escape length is
$X \propto v(p/Z)^{-1/3}$ at $E > 40$ GeV/nucleon.

The second explanation \citep{Ptus06} assumes that the
Iroshnikov-Kraichnan spectrum describes the interstellar MHD
turbulence. It is characterized in particular by a relatively slow
nonlinear cascade of wave from small to large wave numbers. The
resonant wave-particle interaction results in the significant wave
damping on cosmic rays and termination of the cascade at
$k\sim10^{-12}$ cm$^{-1}$ (in contrast to the case of Kolmogorov
cascade which is fast and not noticeably affected by cosmic rays).
The amplitude of short waves is suppressed and the low energy
particles rapidly exit the Galaxy without producing many
secondaries. It explains the peaks in secondary/primary nuclei
ratios at about 1 GeV/nucleon, see Figure~(\ref{B_C}). The
asymptotic behavior of the scape length in this case is $X \propto
v(p/Z)^{-1/2}$ at $E > 3$ GeV/nucleon.

It is clear from Figure~(\ref{B_C}) that the experimental data on
the abundance of secondary nuclei are well described by both
physical models. Also illustrated in Figure~(\ref{B_C}) are the
results of calculations in the plain diffusion model with the
empirical diffusion coefficient of the form $D\propto \beta^{-2}$
and $\beta^{-2}(p/Z)^{0.6}$ below/above $3$ GV chosen to fit the B/C
data (the "unphysical" model).

It must be emphasized that the theoretical description of MHD
turbulence is a complicated and not completely solved problem even
in the case of small-amplitude random fields. Since the mid 1990s,
there has been a renewed interest in understanding of
magnetohydrodynamic turbulence as it applies to the interstellar
magnetic field and density fluctuations \citep{Gol95,Gol97,Gal00}.
\citet{Gol95} exploited anisotropy in MHD turbulence and obtained
Kolmogorov-like spectrum for the energy density of Alfv\'en waves.
The main part of the energy density in this turbulence is
concentrated perpendicular to the local magnetic field wave vectors
$k_{\perp}\approx k$, while the parallel wave numbers are small:
$k_{\parallel}\sim\left[ kW(k) /\left( B_{0}^{2}/4\pi \right)\right]
^{1/2}k_{\perp}$. The cascade is anisotropic with energy confined
within the cone $k_{\parallel}\propto k_{\perp}^{2/3}$. This
turbulence does not significantly scatter cosmic rays. The
distribution of slow magnetosonic waves passively follows that of
Alfv\'en waves \citep{Lit01}. They are not damped on cosmic rays
(because of the property $k_{\perp}\gg k_{\parallel}$) and are
probably responsible for the observed interstellar electron density
fluctuations. The fast magnetosonic waves with the Iroshnikov -
Kraichnan spectrum $W(k)\propto k^{-3/2}$ may have an independent
nonlinear cascade which is isotropic and can efficiently scatter
cosmic rays. These conclusions were supported by numerical
simulations by \citet{Cho02}. This concept of the MHD turbulence
favors the second of the discussed above scenarios where cosmic rays
are scattered by fast magnetosonic waves with the
Iroshnikov-Kraichnan spectrum and suppress the turbulence at large
wave numbers $k \geq 10^{-12}$ cm$^{-1}$.

The knowledge of the diffusion coefficient is absolutely essential
for understanding the nature of the spectrum of galactic cosmic rays
that is determined by the processes of acceleration in the sources
(supernova remnants) and propagation in galactic magnetic fields.
Two specific asymptotic power laws of the diffusion coefficient
$D\propto (p/Z)^{1/3}$ and $D\propto (p/Z)^{1/2}$ at very high
energies together with the observed spectrum approximated by the
power law $J\propto E^{-2.7}$ imply the cosmic ray source spectra
close to $q\propto E^{2.2}$ and $q\propto E^{2.4}$ respectively.

The contemporary modelling of cosmic ray production by supernova
remnants was achieved by \citet{Ptus10}. The spectra of high-energy
protons and nuclei accelerated by supernova remnant shocks were
calculated taking into account magnetic field amplification and
Alfv\'enic drift both upstream and downstream of the shock for
different types of supernova remnants during their evolution. Four
different types of SNRs (Ia, Ib/c, IIP and IIb) with corresponding
burst rates were included in the calculations. The action of cosmic
ray pressure on the shock structure was taken into account in the
calculations.  It was found that the maximum energy of accelerated
particles may reach $5\cdot10^{18}$ eV for Fe ions in Type IIb SNRs.
The steady state spectrum of cosmic rays produced by SNRs in the
Galaxy was calculated with the deduced source spectra of protons and
different kind of nuclei up to Iron. The escape length with a high
energy asymptotic $X \propto (p/Z)^{-0.54}$ determined by
\citet{Jones01} by the accurate fit to B/C data was used in the
calculations. The derived energy spectrum of cosmic rays including
the knee structure around $3\times 10^{15}$ eV is in good agreement
with the spectrum measured at the Earth from low energies to about
$5\times 10^{18}$ eV. This result is strongly in favor of the cosmic
ray galactic diffusion on the MHD turbulence with spectrum close to
$k^{-3/2}$. The Kolmogorov type spectrum $k^{-5/4}$ is unlikely.

The diffusion approximation cannot be used when the diffusion mean
free path $l=3D/v$ exceeds the size of cosmic ray halo $H\approx 4$
kpc that occurs at $E > 2\times 10^{16}Z$ eV if $D\propto
(p/Z)^{1/2}$. Calculations of particle trajectories in galactic
magnetic field are needed to study cosmic ray propagation at these
ultra high energies. At weaker dependence, $D\propto (p/Z)^{1/3}$,
the diffusion approximation breaks at higher energies. The main
limitation in this case is the inefficiency of particle scattering
at $r_{g}k_{\mathrm{L}}>1$ when the diffusion coefficient along the
average magnetic field rapidly rises with energy as $D\sim
vr_{g}^{2}k_{\mathrm{L}}$, see e.g \citet{Top85}. This regime is
realized at $E> Z\times 10^{17}$eV. The direct numerical modelling
of particle motion in the regular and random galactic magnetic
fields \citep{HorTr} confirmed the presented order of magnitude
estimates. The most essential feature of cosmic ray transport at
ultra-high energies above about $10^{17}$eV is the prevalence of the
drift motion across predominantly azimuthal average galactic
magnetic field \citep{ZirTr}. It leads to the inverse dependence of
the exit time from the Galaxy on particle momentum $T\propto
(p/Z)^{-1}$.

\subsection{Instabilities and plasma effects} Cosmic rays not always
can be treated as test particles moving in given regular and random
magnetic fields \citep{Ginz65,wentzel74,cesarsky80,Kuls05}.
The energy density of cosmic rays estimated as $w_{cr}=1-2$ eV
cm$^{-3}$ is approximately equal to the energy density of the
magnetic field and to the energy density of turbulent motions of the
interstellar gas. The presence of relativistic charged particles in
the interstellar medium leads to collective (plasma) effects during
cosmic ray propagation.

A notable example is the cosmic ray streaming instability which
develops when the bulk velocity of cosmic rays exceeds the Alfv\'en
velocity $V_{a}$. The growth rate of MHD waves amplified by
relativistic charged particles with number density $n_{cr}(E)\propto
E^{-\gamma +1}$ at the resonance wave number
$k_{\textrm{res}}=r_g^{-1}(E)$ is

\begin{equation}
\Gamma_{\textrm{cr}}\approx
\Omega_{\textrm{p}}\frac{n_{\textrm{cr}}}{n}\left(\frac{v\delta_{\textrm{cr}}}{(\gamma
+ 2)V_{\textrm{a}}}-1\right),
 \label{eqgrowth}
\end{equation}
where $\Omega_{\textrm{p}}$ is the gyrofrequency of thermal protons,
$\delta_{\textrm{cr}}$ is the amplitude of cosmic ray anisotropy.
Even for small anisotropy, $\delta_{\textrm{cr}}\approx 10^{-3}$,
the instability for galactic cosmic rays with energies $\sim 100$
GeV develops in about $10^{5}$ years, i.e. rather rapidly for the
galactic timescale. The development of the instability leads to
isotropisation of the angular distribution of particles and
turbulence amplification, see \citet{Zweib03,Farm04} and references
cited therein. Plasma effects make the overall picture of cosmic ray
diffusion in the Galaxy more complicated than it was discussed in
Section 3. In principle, the cosmic ray diffusion coefficient should
be calculated selfconsistently with the account taken for the
generation of turbulence by streaming cosmic rays. The examples of
such approach and the corresponding transport equations can be found
in the papers by \citet{skilling75a,skilling75b} and \citet{pzp09}.
The necessity of considering the variety of dissipation processes of
linear and nonlinear wave dissipation in the interstellar medium
adds complexity to the investigation and makes the results strongly
dependent on the interstellar gas parameters: the density, the state
of ionization, the temperature. The wave dissipation makes the
effect of streaming instability not efficient at high energies since
the cosmic ray density is going down with particle energy and the
growth rate Eq.~\ref{eqgrowth} is decreasing correspondingly. One of
the exceptions is the model of galactic wind driven by cosmic rays
in a rotating galaxy developed by \citet{Zirak96} (it is briefly
described at the end of the present Section). Moving predominantly
along the very long spiral magnetic field lines, the cosmic rays
have enough time to amplify the resonant waves even at high energies
of the order 10$^{15}$ eV.

The effect of the streaming instability is more significant in the
vicinity of the sources. The amplification of magnetic field by the
cosmic-ray streaming instability at the shock in supernova remnant
is an integral part of cosmic ray acceleration
\citep{Bell78,be87,md01,Bell04,zp08,mc10,boe11}. The cosmic rays
produce turbulence which selfconsistently determines the cosmic ray
diffusion coefficient in the shock vicinity.

Cosmic rays may induce, in addition to kinetic effects,
significant hydrodynamic effects in the Galaxy. Accounting for
cosmic ray pressure is principally important for the formation of
a halo filled with gas, magnetic field, and relativistic
particles. The equilibrium distribution of the interstellar medium
above the galactic plane in the gravitational field of stars is
subject to the Parker instability \citep{Park66}. Cosmic rays play
a significant role in the development of this instability. The
instability gives rise to large-scale turbulence and helps sustain
an almost equipartition energy distribution among cosmic rays,
magnetic fields, and turbulent gas motions. The characteristic
time for instability development is $\sim 10^7$ years in the
gaseous galactic disk, and $\sim 10^8$ years in the gas halo.
\citet{Park92} showed that magnetic arches and loops appearing
above the galactic disk due to the action of cosmic rays are
necessary for $\alpha\omega$ dynamo to operate, which is primary
mechanism of magnetic field generation in the Galaxy. A numerical
model of the magneto-hydrodynamical dynamo, driven by cosmic rays in
the interstellar medium was developed by \citet[][]{hanaszea04}. The
cosmic rays in the model were accelerated in randomly occurring
supernova remnants. The cosmic ray propagation was accounted  with
the diffusion-advection equation supplementing the MHD equations
\citep[see also][]{hanaszea06}. The other essential elements of the
model are: vertical gravity of the disk, differential rotation and
resistivity leading to reconnection of magnetic field lines. They
obtain amplification of the large-scale magnetic field on a
timescale of galactic rotation. The model includes the ideas of a
fast, cosmic ray driven, galactic dynamo proposed by \citet{Park92}.
The authors find that both the resistivity and the introduced  SN
activity enhance the efficiency of the cosmic-ray dynamo. The
timescale of magnetic field amplification in this model was as short
as 140 Myr \citep{hanaszea06}. The models assumed some CR diffusion
prescriptions. A linear study of the effect of the anisotropic CR
diffusion with very different but finite parallel
($\kappa_{\parallel}$) and transverse diffusion coefficients was
performed by \citet{ryuea03}. It has been shown that the finiteness
of parallel diffusion slows down the development of the Parker
instability. However, with a realistic value of $\kappa_{\parallel}
= 3\times10^{28}$ cm$^2$ s$^{-1}$ in the ISM, the maximum growth
rate is smaller by only a couple percent than that for
$\kappa_{\parallel}\rightarrow \infty$, and the range of unstable
wavenumbers remains the same. The inclusion of perpendicular
diffusion with $\kappa_{\parallel}=0.02 \kappa_{\perp}$ does not
change the growth rate noticeably. It should be noted however, that
the turbulent advection  due to magnetic fluctuations of scales
larger than the CR mean free path would make the diffusion
coefficient nearly isotropic on scales interesting for the Parker
instability \citep[see][]{bt93}. A non-linear development of Parker
instability  was simulated by \citet{knk04}. They stated that the
growth rate of the instability is larger in the models with smaller
diffusion coefficient only in the early linear stage. On the later
stages the growth rate becomes smaller when compared to that of the
large diffusion coefficient model. The growth of instability is
impeded by the CR pressure gradient force interfering with the
falling motion of the matter in the small- $\kappa_{\parallel}$
model, while the magnetic loop can grow up to larger scales in the
models with large $\kappa_{\parallel}$.

\begin{figure}
\includegraphics[width=0.70\textwidth]{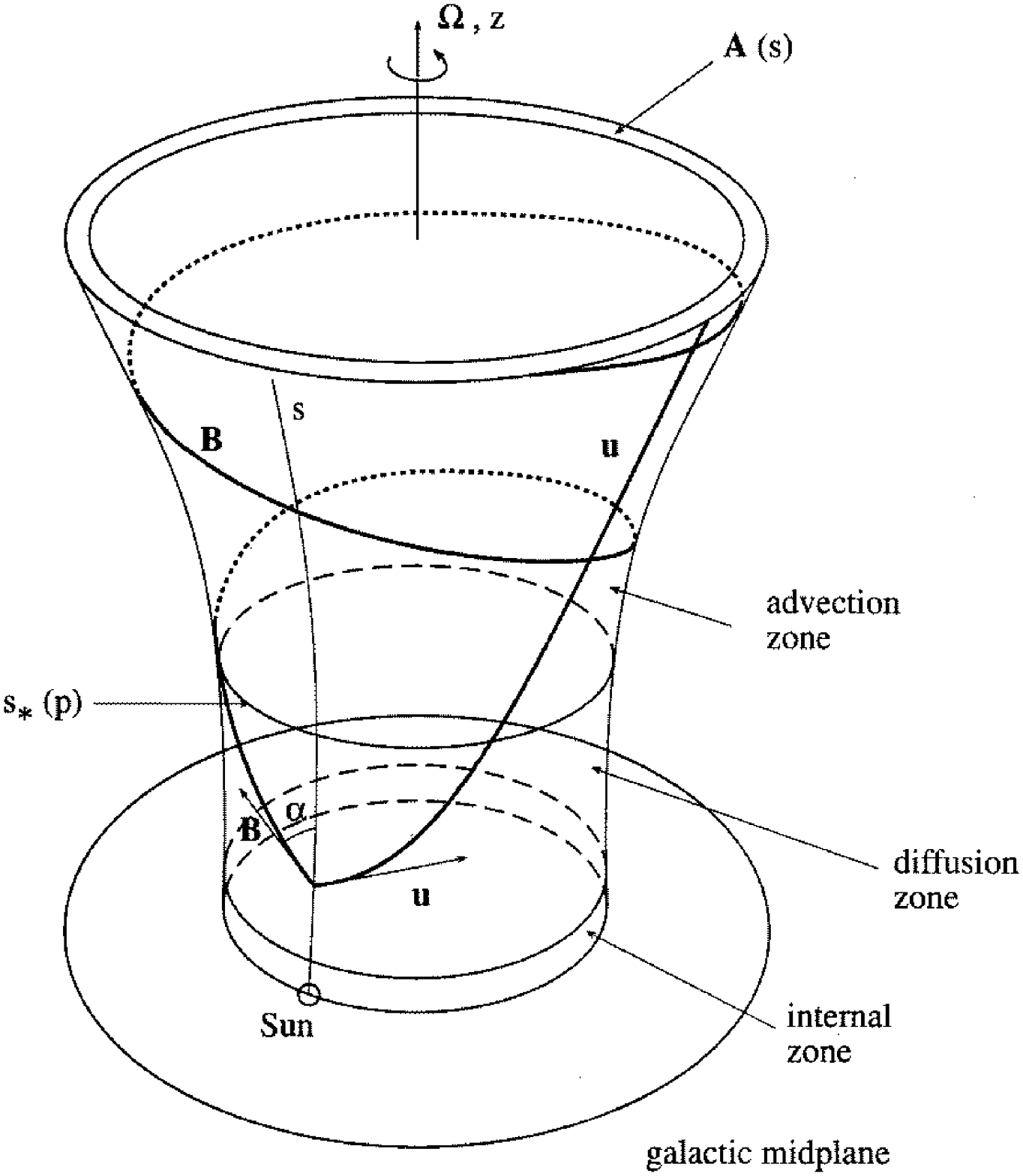}
\caption{The structure of the galactic wind flow driven by cosmic
rays \citep{Ptus97}. The large-scale magnetic field lines, the flow
streamlines and the trajectories of cosmic ray bulk motion from the
Galaxy are confined in the same flux tubes. The cylindrical flux
tube with the cross section $\mathbf{A}(s)$ originating at the Sun
location is shown (here $s$ is the coordinate along the flux tube).
Diffusion mainly determines the cosmic-ray transport in the
diffusion zone (the analog of the boundary layer), whereas
convection mainly makes it in the advection zone. The boundary
between two zones $s_{*}(p)$ is moving up with energy of cosmic-ray
particles.} \label{Wind}
\end{figure}

It is possible that the gas in the galactic halo is not in static
equilibrium but is involved in large-scale convective motion - the
galactic wind. The data on galactic soft X-ray emission suggest that
a wind exists in our Galaxy, see \citet{Ever07}.
It can be supported by cosmic ray pressure \citep{Ipav75,Breit91}.
A model was constructed by
\citet{Zirak96} where cosmic rays, after leaving the sources
(supernova remnants), determine the wind outflow in the rotating
Galaxy with a frozen magnetic field.  see Figure~(\ref{Wind}). Here,
the streaming instability of cosmic rays exiting the Galaxy along
the spiral magnetic field lines leads to the MHD turbulence
generation, that self-consistently determines the
diffusion-convection transport of relativistic particles. The level
of turbulence is regulated by the nonlinear Landau damping on
thermal ions. The outflow velocity is $\sim 30$ km s$^{-1}$ at a
distance $\sim 20$ kpc, and the speeds up to a velocity of $\sim
400$ km s$^{-1}$ several hundred kpc away. The external pressure of
the intergalactic gas produces a termination shock at a distance of
$\sim 300$ kpc. In this model, the diffusion coefficient of cosmic
ray is not given independently and is self-consistently calculated,
being dependent on the power of sources and the spectrum of
accelerated particles \citep{Ptus97}.
Remarkably, the obtained transport coefficients and other model
parameters are consistent with the empirical diffusion-convection
model, e.g. \citet{Bloem93}, for cosmic ray propagation in the
model with galactic wind. It should be emphasized that the existence
of a galactic wind, either thermal or driven by cosmic rays, remains
hypothetical. However, \citet{everettea08} found that the observed
galactic soft X-ray emission can be better explained by a wind than
by the static gas models and the cosmic ray pressure is essential to
drive the wind.

\section{The highest energy cosmic rays}
The "Hillas diagram" discussed above predicts that the compact
sources like gamma ray bursts (GRB) \citep[see e.g.][for a
review]{pm06} are potential sources of UHECRs
\citep{GRB1,GRB2,GRB3,waxman04}, as well as the cosmological sources
like active galactic nuclei (AGN), radio galaxies \citep[see
e.g.][]{longair10,blasi06}, and clusters of galaxies
\citep[e.g.][]{nma95}. The observed widespread warm-hot
intergalactic gas was likely heated by large scale cosmological
shocks. The shocks in the Large Scale Structure formation scenario
are driven by gravitationally accelerated flows
\citep[e.g.][]{kangea05}. Non-thermal emission observed from
clusters of galaxies indicate the presence of accelerated particles
in scales larger the galactic halo sizes
\citep[e.g.][]{ferrariea08}. Large-scale accretion shocks in the
cluster of galaxies and more generally cosmological shocks in Large
Scale Structure can accelerate particles to energies above 10$^{19}$
eV. Namely, \citet{nma95} argued that cosmological shocks can be
good sites for UHECR acceleration if there is an intergalactic field
of order 10$^{-9}$ G and even more important if microgauss regime
fields can be self-generated in shocks accelerating particles. The
maximal energies cosmic rays accelerated in the large scale shocks
like the cluster accretion shocks or hot spots in radio galaxies can
reach $\sim$ 10$^{19}$ eV regime only if indeed microgauss fields
can be produced \citep[][]{nma95}. Current cosmological shock
simulations \citep[e.g.][]{bdd08,dbd08} and the non-thermal emission
models \citep[e.g.][]{kushnirea09} are in favor of the reality of
microgauss regime magnetic fields in clusters of galaxies. On the
other hand the losses caused by interactions of the UHECRs with
photons of the Cosmic Microwave Background Radiation could limit the
maximal energies of protons accelerated in clusters of galaxies
\citep[see e.g.][]{vannoniea09}.

\subsection{Ultra-high energy cosmic rays as probes of magnetic fields}

At the highest energies, the gyroradius of the cosmic rays becomes
extremely large, especially in the intergalactic space where the
magnetic fields are expected to be much lower than in the Galaxy.
Their mode of propagation thus changes from being diffusive to being
asymptotically rectilinear, so at extreme energies it is realistic
to think of the possibility of a cosmic ray astronomy, with the
sources of ultra-high energy cosmic rays (UHECRs) being identifiable
directly.  For example a proton of energy $6\,10^{19}$~eV will be
deflected by 1$^{\circ}$ to 5$^{\circ}$ in the galactic magnetic
field depending upon the direction and length of the trajectory.

On the one hand, being able to actually \emph{see} the UHECR sources would be a remarkable breakthrough:
 i) it would (at least partially) solve the mystery of their origin,
 ii) it would put strong constraints on the acceleration mechanism and its efficiency,
 teaching us a lot about the physics of particle acceleration in the universe in its most extreme
 manifestations, and iii) it would offer an important piece of information, complementary to
 that gathered from multi-wavelengths observations of photons, to guide us
in the astrophysical modeling of
 the sources. This would indeed enhance the power of current multi-messenger analysis, which
are mostly relying on upper limits.

On the other hand, if we could be sure to know the source of (at
least a set of) UHECRs, we could use these particles as messengers,
not only to gather new information about the sources, but also to
probe the intervening magnetic fields, by studying the deflections
they suffered along their journey to the Earth.

In the following, we review the current status of UHECR observations, and discuss some of their implications.
We also discuss why a new generation of detectors can be expected to lead to the above-mentioned breakthrough,
and what can already be learned from the current data, including from the fact that its statistics is still
insufficient. But to understand the data, we first need to understand the transport of UHECRs, and see how
it can modify their energy spectrum, their composition and their arrival directions.

\subsection{UHECR transport: energy, mass and direction}

The key ingredient of UHECR transport in the extragalactic space is
the so-called GZK effect, which refers to the interactions of the
high-energy cosmic rays with the background radiation fields in the
universe~\citep{Gr,ZatsK}. In the case of high-energy protons, the
dominant interactions are with the photons of the cosmological
microwave background (CMB), producing electron-positron pairs and/or
pions as soon as the proton energy is high enough for the CMB
photons to exceed the reaction threshold, when boosted in the proton
rest frame. These interactions lead to strong energy losses,
especially above $\sim 6\,10^{19}$~eV, when photopion production
becomes important, which significantly limits the distance UHE
protons can propagate in the universe~\citep[see e.g.][for a
review]{BhaSig00}. This translates into an effective horizon, from
beyond which extragalactic sources cannot contribute to the observed
UHE protons, and this in turn implies an expected strong suppression
of the flux, referred to as the GZK cutoff. At lower energy, energy
losses due to photopair-production on the CMB photons also modify
the spectrum of UHE protons, producing a dip around $3\,10^{18}$~eV,
where the ankle is observed in the overall CR
spectrum~\citep{Blumenthal70,BerGri88}. Another simple
interpretation of the ankle, however, is that it marks the
transition from galactic to extragalactic cosmic rays.

As already noted in the original work of \citet{Gr} and
\cite{ZatsK}, a similar effect holds for heavier nuclei, which can
be photodissociated by CMB and/or infrared photons as they propagate
through the intergalactic space. The propagation of
ultra-high-energy nuclei in the intergalactic space was studied in
detail by \citet{PSB76}, and recently revisited with updated and
additional cross-sections~\citep{Khan+05,Allard+05}. A key result is
that photodissociation is very efficient at ultra-high-energies, and
lead to: i) a modification of the composition of UHECRs as they
propagate from their sources to the Earth, as well as ii) energy
losses associated with the loss of nucleons (although pair
production also plays a role at lower energy, to a  first
approximation the nuclei essentially keep a constant Lorentz
factor).

The energy threshold for the photodissociation of UHE nuclei depends on the energy of the background
photons, as well as on the mass of the nuclei. At the threshold, the dominant process is the excitation
of the giant dipole resonance, involving photons around 10~MeV in the nucleus rest frame.
Photons in the tail of the CMB distribution reach this energy for nuclei with a Lorentz factor
of the order of $\Gamma \ga 10^{9}$, and energies $E \sim A\times\Gamma m_{\mathrm{p}}c^{2}$, where $A$ is the mass number of the nucleus. Thus, heavier nuclei can survive photodissociation up to higher energies. By a striking coincidence of Nature, although the processes involved are very different, and the cross section thresholds and sizes are also different, the energy loss length for UHE protons and for Fe nuclei are essentially the same in the intergalactic space, so both species have a very similar horizon structure. Therefore, their elemental GZK cutoff leads to the same spectral shape, and indeed the energy distribution of UHECRs can be understood just as well in terms of pure proton sources or of pure Fe sources (or a combination of both, provided the source spectral index and/or the evolution of the source luminosity as a function of redshift is adjusted correspondingly). In principle, intermediate mass nuclei may also be accelerated in the same UHE sources. However, at a given energy, their Lorentz factor is smaller than the Lorentz factor of Fe nuclei, so they see higher energy photons and get dissociated more rapidly. As a consequence, the very highest energy
cosmic rays are likely to be mostly protons and/or Fe nuclei (heavier nuclei would also survive, but \emph{a priori} they are much less abundant in the sources).

The studies on the propagation of UHE nuclei by
\citet{Allard+05,Allard+07} also drew a renewed attention to the
question of the galactic/extragalactic transition, whose energy
scale turns out to depend on the assumed UHECR source composition.
Schematically, a pure proton composition (at the sources) would
allow one to fit the high energy spectrum down to below
$10^{18}$~eV, assuming a steep source spectrum (in $E^{-\alpha}$,
with $\alpha \sim 2.6$ or 2.7, depending on the assumptions on the
cosmological source evolution), and would then imply a
galactic/extragalactic transition at a few times $10^{17}$~eV.
Conversely, a mixed source composition (e.g. similar to that of low
energy cosmic-rays in the Galaxy) would allow one to fit the energy
spectrum with a source spectral index $\alpha \sim 2.3$, and imply
that the extragalactic component overcomes the galactic one at the
ankle~\citep{Allard+05}. We note in passing that this question is
also related to the question of the galactic magnetic fields, which
must be able to confine the cosmic rays up to the transition energy.

Any information about the UHECR composition would thus have
important implications for the general phenomenology of cosmic rays,
including low energy ones. Unfortunately, a corollary of the above
remarks is that the overall UHE spectrum cannot be used, on its own,
to constrain the source composition of the UHECRs. A direct
measurement of the mass of UHE particles is thus very important, and
although this is experimentally difficult, noticeable progress has
been made recently (see below). However, it is fair to say that the
experimental situation is not clear yet, and since the measurements
also rely on hadronic physics at energies well beyond the energy
accessible through direct experiments, it is not clear how
conclusively this kind of measurements can resolve the issue in the
near future.

In this context, two other aspects of UHECR transport can play a
role, which are worth mentioning here since they also involve
magnetic fields. First, the interactions associated with the
above-mentioned GZK effect(s) produce secondary particles, including
the so-called cosmogenic neutrinos and the electromagnetic cascades
initiated by secondary gamma-rays (from pion decay) or electrons and
positrons. These cascades involve the usual mechanisms of
synchrotron radiation, pair production and inverse Compton
interactions~\citep[see e.g., in the context of UHECR
propagation,][]{BhaSig00}. These secondary particles can in
principle be detected (or their fluxes constrained). In the case of
secondary neutrinos, the expected flux and energy distribution
strongly depends on assumptions about the phenomenology of the UHECR
sources~\citep[for a recent update, including scenarios with a mixed
composition, see][and references therein]{Allard+06,Kotera+10}. It
should also be noted that secondary particles produced in the
acceleration site may be detected or constrained, and thus provide
additional, multi-messenger information about the UHECR sources and
their environment, as well as about the acceleration mechanism
itself. Finally, extragalactic magnetic fields can also be
constrained by the observation of TeV gamma-ray halos around
potential acceleration sites. Recently \citet{tvn11} discussed the
constraints on the strength of extragalactic magnetic field derived
from the simultaneous detections of TeV emitting blazars in GeV
photons by the \emph{Fermi} telescope. The measured GeV flux was
found to be lower than that calculated in the cascade model under
the assumption of zero magnetic field. Assuming that the reason for
the suppression of the cascade component is the extended nature of
the cascade emission, the authors concluded that the extragalactic
magnetic field above 10$^{-15}$ G of correlation length of ~1 Mpc is
consistent with the data.

The second aspect, directly associated with magnetic fields, is the
deflection of UHE particles between their sources and the Earth. If
the source of each individual cosmic ray was known, the angular
distance between the source and the arrival direction would give a
very precious measurement of the deflection integrated along the
path of the cosmic ray. Since the deflection, in a ballistic or semi
ballistic regime, is proportional to the particle charge, an
analysis of the deflection patterns would provide information about
the UHECR charges and/or about the intervening magnetic fields.
Although the identification of sources apparently remains out of
reach of the current detectors, we shall mention below what the
current observational situation can nevertheless teach us, and why
this situation is expected to change with a ten times larger
statistics at the highest energies. It should also be mentioned that
additional hints about the composition of UHECRs might come from the
study of their anisotropies, comparing patterns in the arrival
directions at different energies and/or angular scale. The basic
idea is that protons of energy $E$ have the same gyroradius as
heavier nuclei with energy $Z\times E$. Now, in a regime where
energy losses can be neglected in a first approximation (say, below
the GZK energy scale), the angular transport of the different UHE
particles only depend on their gyroradii. Therefore Fe nuclei from a
given source follow exactly the same path as protons at an energy
26~times lower. This has recently been turned into a very
interesting argument by \citet{LemWax09}, relating anisotropy
studies at different energies to composition assertions. Extensive
studies with larger statistics should provide important clues about
this crucial issue in the future. Likewise, at the highest energies,
if the number of sources is low and a few hot spots appear in the
UHECR sky, halos of different angular sizes around the centroid of
the source should appear, corresponding to different nuclear species
(essentially protons and Fe or sub-Fe nuclei, if both are present at
the source).

More generally, it is clear that the study of the UHECR angular
deflections have a lot to teach us, both about the cosmic rays and
about the intervening magnetic fields. The main aspects of this
transport can be summarized as follows. We may distinguish
deflections occurring: i) at or around the sources, ii) in the
intergalactic space, and iii) in the Milky Way. Even though the
sources remain unknown, it is not unlikely that they are embedded in
environments where large scale magnetic fields exist, e.g. in
galactic clusters. These magnetic fields may be able to deflect even
the highest energy CRs significantly, or even confine them for some
time before they escape from the source environment. This can have
an effect of the energy spectrum reaching out in the intergalactic
space, if the confinement time becomes comparable to the energy loss
time. It may also affect the composition of the UHECR leaving the
sources, since, at a given energy, heavier nuclei are more easily
confined. However, this would require effective lengths of several
tens of~Mpc at $\sim 10^{20}$~eV. While such a confinement would
increase the angular size of the source on the sky, turning it, say,
from a localized source to the size of a whole cluster of galaxies
for instance, this should not affect our ability to identify
clusters of events from individual sources on the sky, as long as
the number of sources is limited (as expected at the highest
energies) and their typical angular separation is larger than the
apparent size of the confinement halo. Finally, it is worth noting
that a specificity of UHE cosmic rays, compared to the low-energy,
galactic ones, is that the constancy of the fluxes cannot be assumed
anymore, since the dynamical timescales of the sources can be
shorter than the confinement timescales around them and/or the
differential propagation timescales to the Earth. This may result in
time dependent effects (on time scales of typically thousands to
tens of thousands of years) as to which sources contribute to the
observed flux at different energies, as well as in different mass
ranges, depending on the elapsed time since their activity as UHECR
sources.

Once they leave their sources, UHECRs may also be deflected by
magnetic fields in the intergalactic voids, if their strength is
large enough. This, however, is highly uncertain. Among the
interesting effects that may be associated with such magnetic
fields, the existence of a magnetic horizon has received particular
attention, as it was realized that it could act as an anti-GZK
effect, i.e. essentially a high-pass filter for energetic particles,
preventing lower energy cosmic rays from reaching us from distant
sources, if they wander around in the extragalactic space (due to
magnetic diffusion) for a time longer than the energy loss
timescale, or even the age of the universe
\citep{Parizot04,Lemoine05,AloBer05}. This can have an impact on the
matching between the galactic and extragalactic components, and also
on the cosmic-ray composition in the corresponding energy range: at
a given energy, heavier nuclei diffuse for a longer time than light
ones, and thus extragalactic Fe nuclei should not be able to reach
us from distant galaxies at a energy as low as protons do. It should
be noted, however, that the actual outcome of such propagation
effects on the UHECR spectrum and composition depend on the
granularity of the sources~\citep{Deligny+04,AloBer05}. If the
sources were uniformly distributed in the universe, no effect at all
would be expected, so no definite prediction can be made at our
current stage of ignorance about the sources, the source density and
of course the intergalactic magnetic fields. More recent works have
focused on the propagation of UHECRs in strongly inhomogeneous
extragalactic magnetic~\citep{KotLem08a,KotLem08b}. This is indeed a
much more plausible assumption, under which UHECRs appear to
propagate essentially in straight lines between more localized
interactions with large scale magnetized structures, acting as
scattering centers. As a consequence, potential hot spots in the
UHECR sky may reveal the position of the most nearby scattering
centers rather than that of the most nearby
sources~\citep{KotLem08b}.

\begin{figure*}
\begin{center}
\mbox{\includegraphics[width=0.5\textwidth,angle=0]{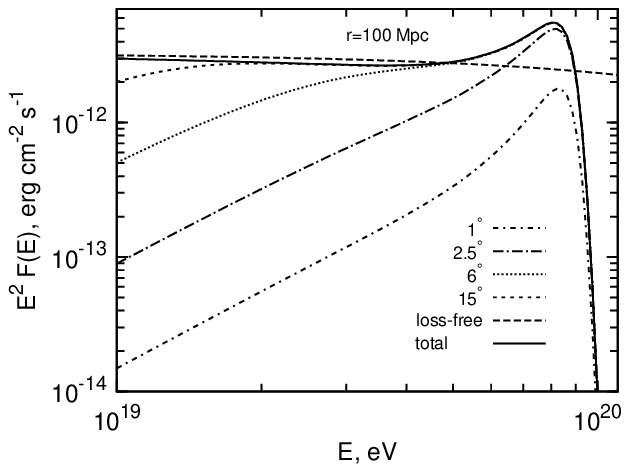}
\includegraphics[width=0.5\textwidth,angle=0]{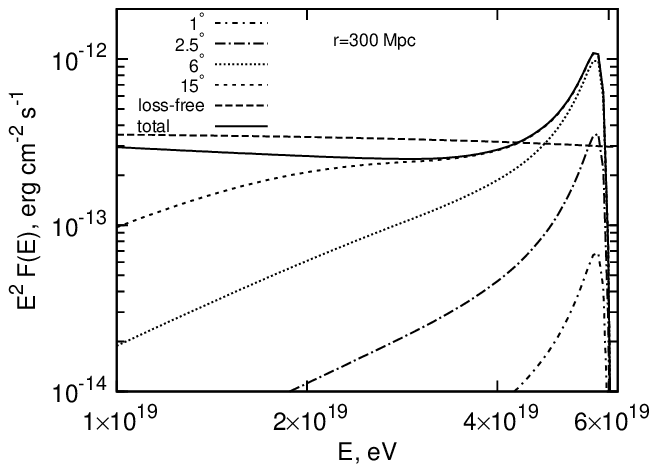}}
\caption{\label{fopr} Energy distributions  of protons observed
within different angles for the source at  distances  $r=100$ Mpc
(left panel) and $r=300$ Mpc (right panel). The energy spectrum of
protons is assumed in the form of power-law with exponential cutoff,
with $\alpha=2$ and $E_0=3 \times 10^{20}$eV;  the rate
$L=10^{44}$erg/s.}
\end{center}
\end{figure*}

Different aspects of  propagation of cosmic rays in IGMF  recently
have been  comprehensively studied \citet{Dolag,Globus,kl08}. The
conclusions of these studies are generally different, basically
because of different assumptions and approaches in the modeling of
the IGMF. The propagations  of protons and accompanying them
gamma-rays and neutrinos recently have been studied by \cite{AKP}
in  small angle approximation limit. In Fig.\ref{fopr}  we  show
the energy distributions of protons within different angles arriving
from a point source located at distances 100 and 300 Mpc. It is
assumed that the source is  injects protons into the intergalactic
medium with magnetic field 1nG and correlation length
$\lambda=1$~Mpc. The upper dashed lines in Fig.\ref{fopr} correspond
to the case when protons propagate in empty space; flux is
determined by the geometrical factor $1/r^2$. The solid line present
the case when the deflections in the magnetic field are ignored.
Comparison of these two curves demonstrates strong dependence of the
energy distribution of protons on the solid angle within which they
are detected; the flux of protons at highest energies is
concentrated along the direction to the source. This is the result
of selective deflection of  of protons depending on their energy. In
addition to this effect there are additional two spectral features
-- a bump and a sagging at lower energies. The  bump preceding the
cutoff is due to strong growth of energy losses at the threshold of
photomeson production that makes particles to be accumulated in this
energy region; the sagging is a consequence of the energy losses due
to the electron-positron pair production.  However,  as it is seen
in Fig.\ref{fopr}, the  maximum due to the
 deflection  is  more  distinct   than the "photoproduction energy loss bump".

Finally, at the other end of the propagation track, the galactic
magnetic fields also deflect the UHECRs, with deflections of the
order of a degree times the charge of the particle, or possibly much
more if an extended magnetized wind surrounds our Galaxy. However,
the exact deflection patterns (both size and direction) depend on
the structure of the magnetic field, which remains highly uncertain.
The regular component should provide a global shift on the sky of
the image of the actual UHECR sources, while the turbulent component
essentially broadens the apparent cosmic-ray source around that
image. Finally, the deflections are expected to be inversely
proportional to energy, which should lead to characteristic
structures in the arrival direction/energy space. Chains of UHECR
events, ordered in decreasing energy away from what would be a
common source, have been sought in the various data sets, with no
conclusive result so far. Obviously, while such structures would be
extremely informative about the source positions as well as about
the magnetic field lines responsible, they can be detected only if
the sources do not overlap (i.e., at the highest energies, but then,
unfortunately, with a limited lever arm in energy), or with much
larger statistics, to allow the detection of these patterns on top
of a background mixing UHECRs from overlapping sources, with a large
enough significance.

In conclusion, the study of UHECR transport in the energy space,
mass space and angular space, makes it clear that the road to a
better understanding of the sources and phenomenology of UHECRs
passes through the joint study of the three fundamental observables
(energy spectrum, composition and arrival directions). It should
also be stressed that the GZK cutoff, while implying a limitation of
the UHECR energies in the universe (independently of the sources)
and strongly reducing the expected flux, is also the best chance we
have to answer the central questions about UHECRs. As the current
data show (see below), no obvious source and/or pattern has appeared
in the sky yet. This implies either that the sources are numerous,
and we have not seen multiplets (i.e. several events coming from the
same astrophysical source), or not strong enough for angular
associations in the sky to be significant, or that the overall
deflections of UHECRs are large, and the sources largely overlap.
The obvious way out is by reducing the number of sources as well as
the deflections. This is precisely what the GZK allows (and actually
imposes). By reducing the horizon scale as the energy increases, the
GZK effect guarantees that fewer and fewer sources contribute at the
highest energies. In addition, the number of sources capable of
accelerating cosmic rays up to a given energy is likely to be a
decreasing function of that energy. In addition, the deflections are
inversely proportional to energy, so by concentrating on the UHECR
sky at $10^{20}$~eV or so, we are bound to face a situation where a
(very) limited number of sources contribute to the observed flux,
with limited deflections and well separated ``centroids'' of the
corresponding hot spots in the UHECR sky. The price to pay for such
a situation, of course, is that the corresponding cosmic ray fluxes
are very low, so further progress is conditioned by the resolution
on an important observational challenge. The target has been set at
the level of a total exposure of the sky of several $10^{5}$ up to
$10^{6}$~km~sr~yr~\citep[see the recent white papers on
UHECRs:][]{Olinto+09}, i.e. more than an order of magnitude larger
than what is currently available.

\subsection{UHECR Observations}

The highest-energy particles are so rare that they are detectable only by
means of the giant cascades or extensive air showers they create in the atmosphere.
Details of how these extensive air showers are observed and of how the parameters of importance
are measured can be found in reviews \citep[e.g.][]{NagWat00}.  The results that are most
relevant to the topic of this chapter have been obtained with
the Pierre Auger Observatory \citep{AugerNIMA,AugerHybrid07}
in Argentina and with the HiRes instrument \citep{HiResNIM02} operated in the Dugway Desert, Utah, USA.
The HiRes device uses telescopes comprising mirrors and photomultipliers to detect the fluorescence
radiation emitted from the excitation of atmospheric nitrogen by the shower particles passing through
the atmosphere.  The Auger Observatory has similar telescopes but in addition has an array of
water-Cherenkov detectors deployed over 3000~km$^{2}$ to measure the particles in the showers at ground level.
  The fluorescence telescopes are used to obtain a measure of how the cascade grows and decays
  in the atmosphere while the water-Cherenkov detectors are used to obtain the pattern of the distribution of shower particles on the ground from which a shower size, closely proportional to the primary energy, is obtained.  With both devices the direction of the incoming cosmic ray is measured with an accuracy of $\sim 1^{\circ}$ using GPS-based timing.  

A 21$^{\mathrm{st}}$ century innovation has been to build instruments combining the fluorescence and the
surface detector techniques. This combination has allowed what is known as the hybrid method to be developed in which the time of the shower front traversing one of the detectors on the ground is combined with the times of arrival of light at the photomultipliers of the fluorescence telescopes, to enhance significantly the precision with which the direction, energy and depth of maximum of the showers can be determined.  The fluorescence technique has the unique advantage of enabling the energy of a shower to be found without resorting to assumptions about hadronic physics.  This makes it possible to calibrate the shower size found with the surface array, 
and so exploit the near 100\% on-time of a surface array.  Two
implementations of the hybrid approach are now operating: the Pierre
Auger Observatory in Argentina and the Telescope Array in
Utah~\citep{TA05}.  Data from the Telescope Array are not yet
available and the exposure from the Auger Observatory (early 2010)
is about 10 times greater.

\subsection{Status of measurements of the energy spectrum}

An energy spectrum based on an exposure of 12,790~km$^{2}$~sr~yr has recently been reported by the Auger Collaboration~\citep{AugerSpectrum10}.  This exposure is a factor 4 higher than achieved with the HiRes detector at the highest energies and even greater than this at lower energies.  When deriving the primary energy from fluorescence measurements, an estimate of the missing energy carried into the ground by hadrons, muons and neutrinos must be made based on assumptions about the mass of cosmic rays and of the hadronic model.  For a primary beam that is a 50/50 mixture of protons and iron, it has been found from simulations of showers with the QGSJET01 model of hadronic interactions that the energy found from the fluorescence signal should be increased by 10\% \citep{Barbosa+04}. The systematic uncertainties on the energy scale sum to 22\% with the largest coming from the absolute fluorescence yield (14\%), the absolute calibration of the fluorescence telescopes (9\%) and that due to the reconstruction method of the longitudinal shower profile (10\%).  Efforts are underway to reduce these uncertainties.

\begin{figure*}
\centerline{\includegraphics[width=0.75\textwidth]{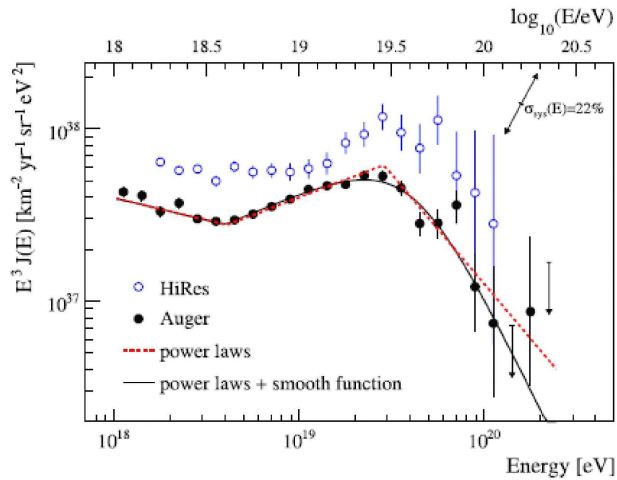}}
\caption{The Auger energy spectrum.  The spectrum is fitted with two functions
[see \citet{AugerSpectrum10} for details] and the data are compared
with the stereo spectrum of HiRes~\citep{HiResSpectrum09}.  The uncertainty of the
flux scaled by E$^{3}$ arising from the systematic uncertainty in the energy scale of 22\%
is indicated.} \label{fig:AugerSpectrum}
\end{figure*}
The Auger spectrum reported in \citet{AugerSpectrum10}, and based on over 36,000 events, is shown in Fig.~\ref{fig:AugerSpectrum} together with the final measurement from the HiRes instrument~\citep{HiResSpectrum09}. It is customary to represent parts of the spectrum with a differential form, $J = kE^{-\gamma}$.  It is seen that there is clear evidence of a flattening of the slope at an energy of $\log E (\mathrm{eV}) = (18.61 \pm 0.01)$ where $\gamma$ changes from ($3.26 \pm 0.04$) to ($2.59 \pm 0.02$).  At $\log E(\mathrm{eV}) = (19.46 \pm 0.03)$ the slope increases very sharply to ($4.3 \pm 0.2$).  This suppression of the flux is what would be expected because of the GZK-effect and is significant at the level of 20 standard deviations.

An intriguing question is whether the slopes in the region of the suppression are different in the Northern and Southern Hemispheres.  This would be expected if the source distribution was different in the two hemispheres with the region with sources at a closer average distance being expected to show a less suppressed slope, assuming that the mean injection spectra are the same.  However the present measurements of ($5.5 \pm 1.8$, HiRes) and ($4.3 \pm 0.2$, Auger) are not significantly different and as the HiRes instrument is not longer operating the possibility of any difference will need to be explored with future instruments.

In any case, one may expect that the  flux at the very highest
energies, namely \emph{in the GZK cutoff region} itself, will be
dominated by only a few sources within the horizon. For this reason,
the assumption of a uniform source distribution usually made to
produce the synthetic spectra used to fit the observed spectrum
should not be valid anymore. Even with infinite statistics, the
overall spectrum around $10^{20}$~eV will not be of much interest,
as it will be shaped mostly by the most luminous sources that just
happen to be here or there in the local sky -- in other words,
\emph{cosmic variance} is the dominant factor in the region of the
spectrum.

The flux above $10^{20}$~eV, as estimated from the 3 Auger events
and 1 HiRes event, is $(2.4^{+1.9}_{-1.1}) \times
10^{-4}\,\mathrm{km}^{-2}\mathrm{sr}^{-1}\mathrm{yr}^{-1}$ or about
1 per square kilometre per millennium.  It is really quite
remarkable that the energy of such rare events can be measured to
within about 20\% and indeed estimates of the energies of particle
initiating showers at the highest energies are more certain than at
$10^{15}$~eV where there has to be greater reliance on models of
hadronic physics.

\subsection{Current status of measurements of the arrival direction distribution}

From a quick look at the first data recorded with the Auger
Observatory at the highest energies it was evident that no strong
sources stood out.  It is not clear what classes of objects are
likely to provide the necessary acceleration sites so a well-defined
search strategy to look for signals was adopted.  Active galactic
nuclei, radio-galaxy lobes and gamma-ray bursts have all been
suggested as possibilities and if these are anisotropically
distributed and are not too numerous within the distance associated
with the GZK-suppression then an anisotropy might be expected.
However there is no prediction for source features as compelling as
was that of the GZK-effect for the spectral shape. Faced with this
dilemma the Auger Collaboration made the decision to search for
correlations on different angular scales with objects in a
particular catalogue using energy and distance as additional
parameters.  The catalogue chosen was the \citet{VCV06} catalogue
(VCV) of quasars and active galactic nuclei.  Using data recorded
between 1 January 2004 and 26 May 2006 a scan was made for a minimum
in the probability $P$ for a set of $N$ events expected from an
isotropic flux to contain $k$ or more events at a maximum angular
distance $\Psi$ from any object in the VCV catalogue.  $P$ is given
by the cumulative binomial distribution
$\sum_{j=k}^{N}C_{j}^{N}p^{j}(1-p)^{N-j}$, where $p$ is the fraction
of the sky (weighted by exposure) defined by the region at angular
separation less than $\psi$ from a selected source. The parameter
space was defined by the angular separation $\psi$, the maximum
red-shift $z_{\max}$ and the threshold energy $E_{\mathrm{th}}$.  A
minimum in $P$ was found with $\psi = 3.1^{\circ}$, $z_{\max} =
0.018$ ($\sim 120$~Mpc) and $E_{\mathrm{th}} = 5.6\times10^{19}$~eV.
With these values, 12 events among 15 correlated with AGNs in the
catalogue, whereas only 3.2 would have been expected by chance.
These numbers were found after a search and so no probability could
be attached to the significance of the association.  However the
observation led to the definition of a test to validate the result
with an independent data set using these parameters a priori.
Details of the test are given in
\citet{AugerScience07,AugerCorrelationPaper08}.  The test was
applied to data collected between 27 May 2006 and 31 August 2007
using exactly the same reconstruction algorithms as for the initial
period.  In this independent data set, which had almost exactly the
same exposure, 8 of 13 events above the same energy threshold were
found to correlate with events in the VCV catalogue.  The
probability of this occurring by chance is $1.7\times10^{-3}$ and
the result was reported~\citep{AugerScience07}.  This correlation
corresponds to an association of ($69_{-13}^{+11}$)\%.

\begin{figure*}
\centerline{\includegraphics[width=0.75\textwidth]{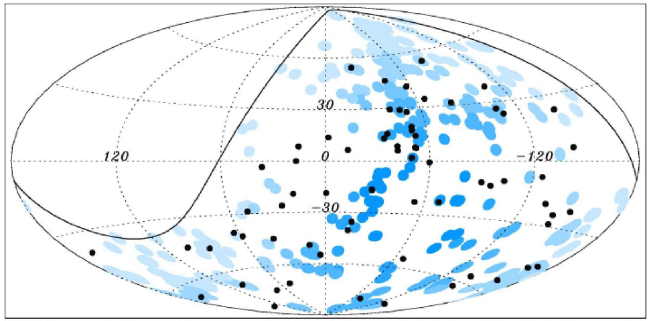}}
\caption{The directions, in galactic coordinates, of 69 events above
$5\times10^{19}$~eV recorded at the Pierre Auger Observatory up to
31 December 2009~\citep{AugerCorrelationUpdate10}.  The events are
shown as black dots on an Aitoff-Hammer projection with the galactic
plane across the centre of the diagram. The shaded circles of
3.1$^{\circ}$ are centered at the positions of 318 AGNs in the VCV
catalogue that lie within 75 Mpc and are within the field of view of
the Observatory. The darker shading corresponds to greater relative
exposure: the exposure weighted-fraction covers 21\% of the sky}
\label{fig:AugerAnisotropy}
\end{figure*}

The Auger Collaboration has continued to take data and as at 31
December 2009 a total of 69 events above the threshold energy have
been recorded~\citep{AugerCorrelationUpdate10}.  The correlation
with objects in the VCV catalogue, excluding the 15 events used to
define the correlation parameters, is now ($38 \pm 6$)\% with 21
events correlated.  The fraction expected from an isotropic
distribution of event directions is 0.21.  From an isotropic flux
the cumulative probability of such a correlation is $P = 0.003$.  A
sky map using an Aitoff-Hammer projection is shown in
Fig.~\ref{fig:AugerAnisotropy}.  With the current estimate of the
correlation at 38\%, a 5 sigma signal ($P < 3\times10^{-5}$) will
require 110 events which should be obtained with about 2 further
years of operation.

Using a cross-correlation study the Auger Collaboration have also
noted that  there is an excess of events from the region around
Centaurus
A~\citep{AugerScience07,AugerCorrelationPaper08,AugerCorrelationUpdate10}
with 13 of the arrival directions within 18$^{\circ}$ of Centaurus A
forming 6 pairs separated by less than 4$^{\circ}$ and 28 pairs
separated by 11$^{\circ}$.  Clearly this is a region of sky that
will be studied carefully but as the search was made {\it a
posteriori} no statistical significance can be attached to the
observation.  No similar object lies as close to us in the Northern
Hemisphere.

The HiRes collaboration does not find a similarly correlation of their events with AGNs in the VCV
catalogue~\citep{HiResAGN08}.  They find that 2 events out of 13 are correlated with VCV objects: this compares with 5 that would be expected assuming that the correlation is really 38\%.  The difference between 2 and 5 does not rule out a 38\% correlation in the northern hemisphere, the part of the sky dominantly observed with the HiRes detector.  It may be that the source distributions are different in the two parts of the sky and, with the steeply-falling spectrum, there may be a small difference in the definition of the energy threshold.  The situation is inconclusive and will not be clarified until new Observatories are built.

However that may be, keeping in mind that the excess of correlation
observed by Auger does not necessarily imply that the AGNs of the
catalog are related with the sources, and may simply reflect the
accidental correlation between these AGNs and the actual sources in
that part of the sky (convolved with the deflections in the
intervening magnetic fields), one must be prepared to a situation
where other sources in the Northern hemisphere lead to a different
level of correlation with the AGNs of that hemisphere. In other
words, there is no direct conflict between the observations
currently available, and a systematic study of the entire sky with
comparable data sets should tell us more about the origin of the
detected anisotropy, and what it implies concerning the UHECR
sources and deflections.

In the mean time, it is interesting to note that the current data
sets do not reveal obvious multiplets or close correlations with
known sources. This may be due to a very large number of sources,
from which only one event at most has typically been observed so
far, or to relatively large deflections, causing the sources to
overlap in the sky. In both cases, the observational solution
consists in increasing the statistics at high energy, where the
deflections are reduced and the horizon scale becomes smaller, so
that only a few sources can significantly contribute. The fact that
some anisotropy is already detectable gives strong confidence that
the expected breakthrough is within reach of the next generation
detectors, provided they can gather a statistics of the order of a
few $10^{5}\,\mathrm{km}^{2}\,\mathrm{sr}\,\mathrm{yr}$ or
more\citep{Olinto+09}.

From the moment when the first few significant hot spots are
observed in the UHECR sky, a new era will begin, with the study of
individual sources. As indicated above, this should lead to key
parameters such as the source power and spectral shape. The current
statistics does not allow one to confirm whether the excess of
events possibly seen around the location of Centaurus A is related
to the first of such sources, but if it is the case, then the next
generation of detectors should accumulate tens or even hundreds of
events from that or similar sources.

\subsection{Measurement of the chemical composition of the primaries}

To interpret the information that we have on the energy spectrum and
arrival direction pattern of ultra high-energy cosmic rays we need
to know the charge, $Z$, of the particles, ideally on an
event-by-event basis.  Then, for example, if one had identified a
point source, the bending of the particles from it as a function of
energy could be used to distinguish between different models of the
magnetic fields lying between the source and earth.  However it is
very difficult to be sure even of the atomic mass, $A$, as the only
methods that are practical for estimating it require assumptions
about the hadronic interactions at centre-of-mass energies well
above what will be reached with the LHC.

The concept of the most promising method is to compare the average depth of
 shower maximum (called $X_{\max}$
) as a function of energy for protons, iron nuclei and photons with
what is predicted for different hadronic models.  The depth of
maximum can be determined to $\sim (20--30)\,\mathrm{g/cm}^{2}$
using the fluorescence technique.  For a mass composition of a
single species, $X_{\max}$ is expected to increase with energy as
$\gamma$-rays from $\pi^{0}$ decay become more energetic and so the
position of shower maximum is pushed deeper into the atmosphere. The
average $X_{\max}$ for iron nuclei lies below that for protons
because the energy per nucleon of an incoming primary is smaller for
heavier nuclei at a given energy.  The change of $X_{\max}$ with
energy is known as the elongation rate.  In addition to the
elongation rate being a useful indicator of primary mass, a key
parameter is the fluctuation in the depth of shower maximum at a
particular energy.  It is intuitively obvious that as an iron
nucleus is larger than a proton, the cross-section for interaction
will be greater.  Thus fluctuations in the position of shower
maximum also contain information about the primary mass.  However,
although these parameters can be measured rather accurately, the
interpretation of the measurements must rely on models of shower
development.

Work on this problem has recently been reported by the Auger
Collaboration \citep{AugerXmax10} and by the HiRes
Collaboration~\citep{HiResXmax10}.  With the Auger Observatory the
depth of maximum has been determined for over 3700 events of energy
$> 10^{18}$~eV.  Each event has been reconstructed using the hybrid
method: 2\% of them  are stereo events for which two independent
estimates of $X_{\max}$ are made.  Using stereo information it can
be demonstrated that the accuracy of estimation of $X_{\max}$ in a
non-stereo event is $\sim 20\,\mathrm{g/cm}^{-2}$ (about 200 m at
the height of the maximum of a typical shower,
$750\,\mathrm{g/cm}^{-2}$, or 2.5 km above sea-level in the vertical
direction).

\begin{figure*}
\centerline{\includegraphics[width=0.5\textwidth]{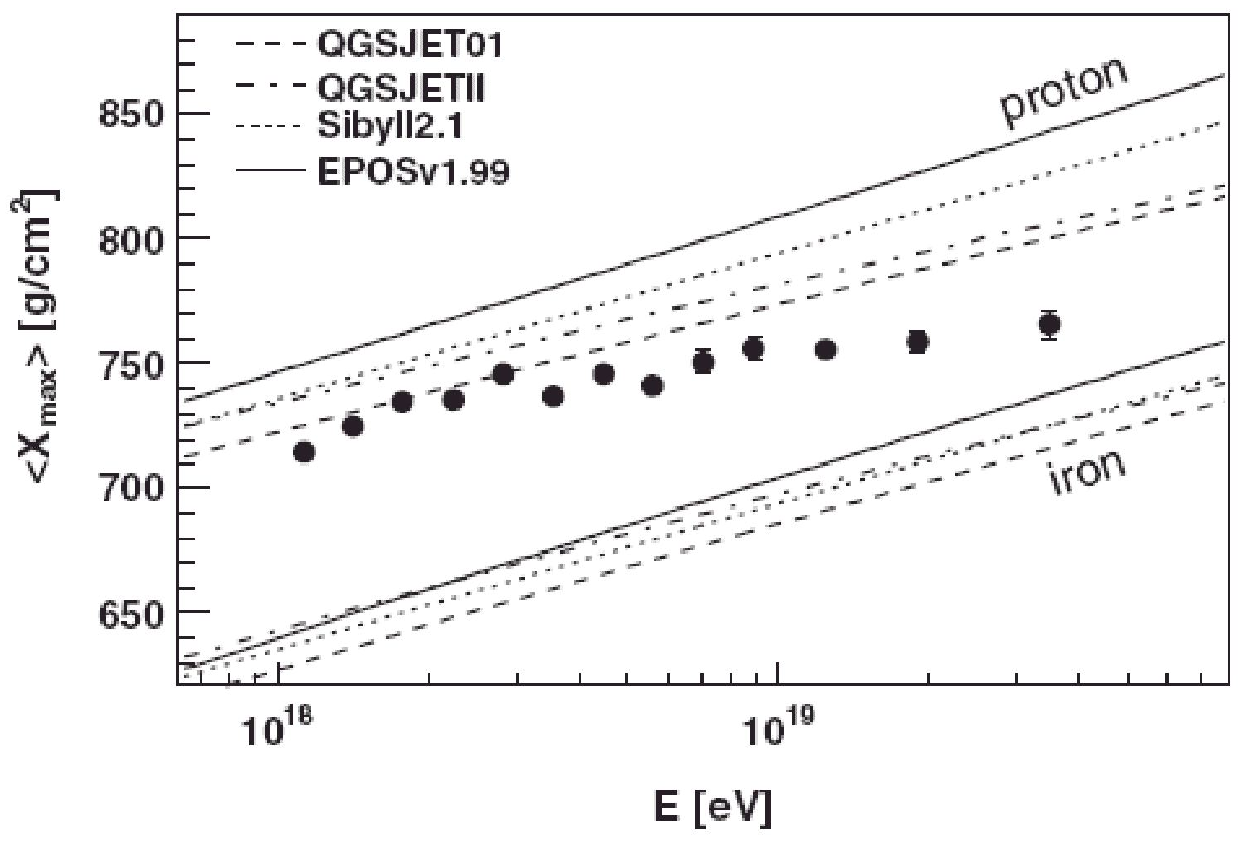}\hfill\includegraphics[width=0.5\textwidth]{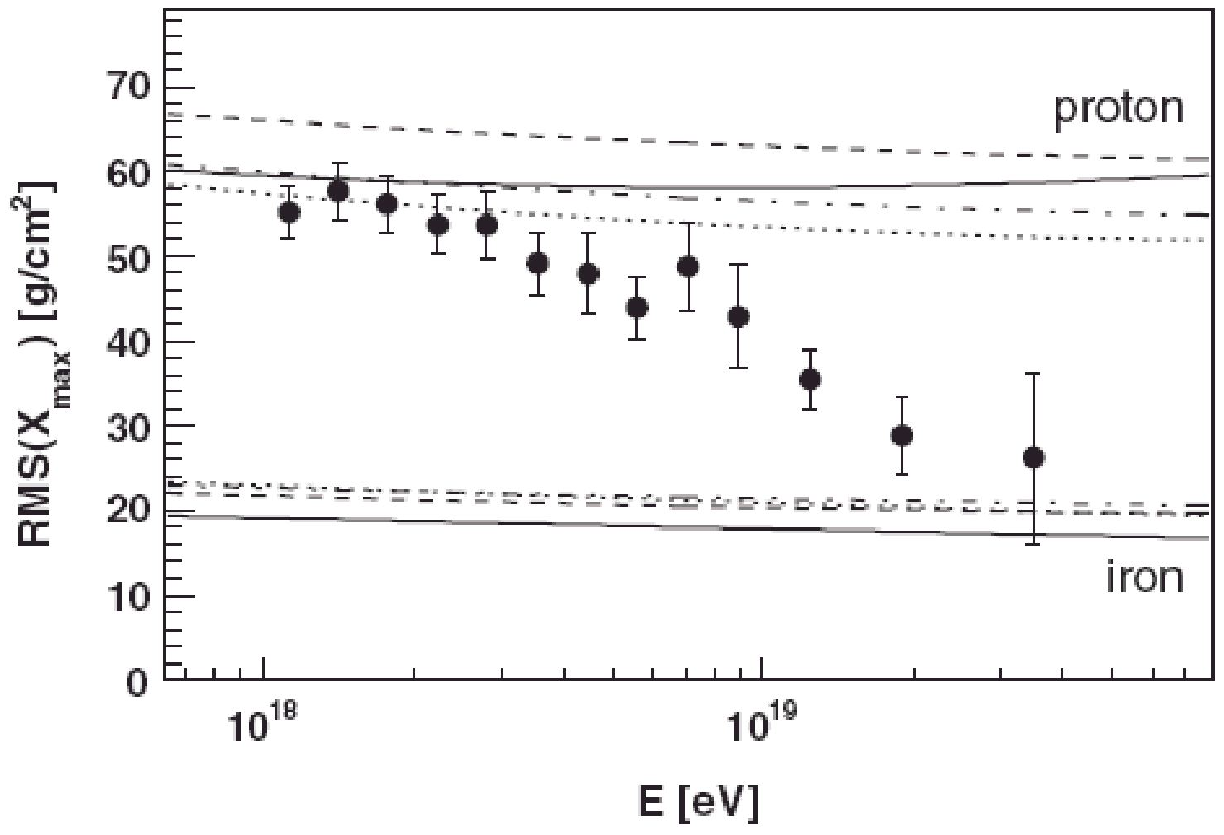}}
\caption{Left: The average measurements of $X_{\max}$ as a function
of energy reported by the Auger Collaboration based on 3754 events
compared with the predictions of 4 different
calculations~\citep[see][for details]{AugerXmax10}. Right: The
measurements of the rms of $X_{\max}$ as a function of energy for
the data of the figure on the left, compared with
calculations~\citep{AugerXmax10}.} \label{fig:AugerXmax}
\end{figure*}

The Auger data on the depth and width of the fluctuation of shower
maximum are shown in Fig.~\ref{fig:AugerXmax}.  The predictions for
proton and iron nuclei of $X_{\max}$ and rms($X_{\max}$) are given
for different models of the hadronic interactions (QGSJET01 etc). It
is clear that if these models are approximately correct then the
mean mass of the cosmic rays is becoming heavier as the energy
increases. However the results from the HiRes collaboration seem to
tell a different story.  They conclude that their data, totaling 815
events, is entirely consistent with proton primaries from $\sim
3\times10^{18}$~eV.  The reasons for the differences between the two
measurements are not understood.  The HiRes result is highly
dependent on Monte Carlo calculations whereas the Auger measurement
makes very little use of such calculations except, as is true for
both measurements, when a comparison is made with the predictions.
The Auger Collaboration have taken precautions in their analysis to
test that the effects observed are not dependent on zenith angle,
demonstrating that deeply penetrating events are not missed, and it
is hard to see how distributions can be made too narrow as a
consequence of the analysis procedures.

\section{Some conclusions and perspectives}
Magnetic fields play a key role in the acceleration, propagation and
radiation of galactic cosmic rays. Current generation of X-ray
\emph{Chandra}, \emph{XMM-Newton} and \emph{Suzaku} and gamma ray
telescopes \emph{H.E.S.S.}, \emph{MAGIC}, \emph{MILAGRO},
\emph{VERITAS} and \emph{Fermi} have provided conclusive evidences
for CR particle acceleration up to the energies at least about 100
TeV in young supernova remnants, starburst regions, radio galaxies
and AGNs. With the coming generation of the focusing hard X-ray
telescopes \emph{Astro-H} \citep{astroH} and \emph{NuSTAR}
\citep{nustar} new possibilities will be opened to study in details
hard X-rays images and spectra that of cosmic ray sources providing
unique information on both particle spectra and fluctuating magnetic
fields in the sources. A project of very large X-ray telescope
\emph{IXO} that is currently under discussion in addition to the
unprecedent sensitivity will be likely supplied with X-ray imaging
polarimetry \citep{ixo11}. Imaging polarimetry is sensitive to
details of magnetic field amplification mechanism in the particle
acceleration sources like supernova remnants and radio galaxies
\citep{bubhk09}. The Cherenkov Telescope Array \emph{CTA} project
under development is the next generation ground-based gamma-ray
instrument, which is supposed to provide at least 5 times better
sensitivity in the current energy domain of about 100 GeV to about
10 TeV and an extension of the accessible energy range well below
100 GeV and to above 100 TeV.

At extremely high energies, $E \sim 10^{20}$ eV, the impact of
galactic and extragalactic magnetic fields on the propagation of
cosmic rays becomes less dramatic, which should result in large and
small scale anisotropies of the cosmic rays. Although the
corresponding fluxes are very low, considerable progress has been
made in the recent years, with the detection of the giant cascades
(so-called extensive air showers) induced by these cosmic rays in
the atmosphere, by large detectors consisting of ground arrays
(AGASA), fluorescence telescopes (HiRes) or hybrid detectors (Auger,
currently in operation).

The extension of cosmic ray studies up to energies of the order of
$10^{20}$~eV is likely to give rise to what is referred to as
``proton astronomy'' (or more generally ``charged particle
astronomy''), where charged cosmic rays are used as astronomical
messengers pointing roughly back to their sources, because of the
reduced deflections in the intervening magnetic fields. Another key
aspect of the cosmic rays phenomenology in this energy range is the
existence of a physical horizon, associated with the energy losses
suffered by the cosmic rays as they interact with the background
radiation fields (notably the cosmological microwave background).

Whether protons or heavier nuclei, particles of such high energies
can arrive only from relatively nearby accelerators, say within
100~Mpc, which decreases dramatically the number of potential
sources capable of accelerating cosmic rays up to $\geq 10^{20}$eV
and contributing to the observed flux. In principle, many weak
sources may be at work, which may prevent one from detecting them
individually, by the accumulation of events in a given direction.
However, even with large source densities, a few most nearby and
brightest sources may be expected to contribute a significant
fraction of the total flux, in a similar way as what is observed at
all photon wavelengths, where a few sources dominate the sky.
Alternatively, there may be only a few sources within the horizon,
especially if one considers the tough requirements associated with
$10^{20}$ eV proton accelerators~\citep{FAetal2002}. This excludes,
in particular, astrophysical objects like ordinary galaxies, unless
these galaxies accelerate the highest energy cosmic rays through
transient events related to compact objects like Gamma Ray Bursts.
Whether we can identify the accelerators of extragalactic cosmic
rays using the highest energy protons is a question which largely
depends on the strength and structure of the large scale IGMF.
However, if the latter are not too large, as seems to indicate the
current UHECR data showing anisotropy above 55~EeV or so, the cosmic
rays from identified sources may be used in turn to probe the
intergalactic magnetic fields in the nearby universe, as well as the
structure and strength of the galactic magnetic field, from the
observed (energy- and charge-dependent) deflections.

The identification of the first cosmic ray sources by astronomical
means, i.e. through hot spots or elongated multiplets with a
characteristic rigidity ordering, may be provided by the next
generation detectors, either on the ground, with the Northern site
of the Pierre Auger Observatory~\citep{AugerNorth2010}, proposed for
construction over 20,000~km$^{2}$, with the proven technique of
Auger South and similar performance, or in space, with the
\emph{JEM-EUSO} mission~\citep{JEMEUSO2010}, proposed to be
installed on the Japanese module of the International Space Station,
for an even larger aperture at the highest energies. This is
expected to lead to a new era of cosmic ray studies, where nearby
sources may be studied individually, and particle acceleration may
be explored in the most challenging regime, up to above $10^{20}$eV.

Finally, we stress that the strong coupling between cosmic ray studies and the understanding of
the magnetic field generation, amplification and structure in the universe calls for an increased exchange
of knowledge between the different communities involved. Important progress is expected in both fields
in the coming years, and any progress in one of the various aspects of the problem will benefit to the others,
and to the whole field of high-energy astrophysics, where energetic particles are responsible for
non-thermal emission at all energies, as well as to the physics of the interstellar medium in general,
in which the magnetic field is a key component.

\begin{acknowledgements}
We thank the referee for the constructive comments. A.M.B. was
supported in part by RBRF grants 09-02-12080, 11-02-00429 by the RAS
Presidium Programm, and the Russian government grant 11.G34.31.0001
to Sankt-Petersburg State Politechnical University. He performed
some of the simulations at the Joint Supercomputing Centre (JSCC
RAS) and the Supercomputing Centre at Ioffe Institute, St.
Petersburg. V.S.P. was supported by RFBR grant 10-02-00110.

\end{acknowledgements}
\bibliographystyle{svjour}
\bibliography{CR_5}




\end{document}